**Analytical model and dynamical phase-field simulation of terahertz transmission across ferroelectrics**

Taorui Chen[1], Bo Wang[3], Yujie Zhu[1], Shihao Zhuang[1], Long-Qing Chen[2], Jia-Mian Hu[1*]

[1]*Department of Materials Science and Engineering, University of Wisconsin-Madison, Madison, WI, 53706, USA*

[2]*Department of Materials Science and Engineering, The Pennsylvania State University, University Park, Pennsylvania, 16802, USA*

[3]*Materials Science Division, Lawrence Livermore National Laboratory, Livermore, CA 94550, USA*

## Abstract

We theoretically investigate the steady-state transmission of continuous terahertz (THz) wave across a freestanding ferroelectric slab. Based on the Landau-Ginzburg-Devonshire theory of ferroelectrics and the coupled equations of motion for polarization and electromagnetic (EM) waves, we derive the analytical expressions of the frequency- and thickness-dependent dielectric susceptibility and transmission coefficient at the thin slab limit in the harmonic excitation regime. When the slab thickness is much smaller than the THz wavelength in the ferroelectric, the analytical predictions agree well with the numerical simulations from a dynamical phase-field model that incorporates the coupled dynamics of strain, polarization, and EM wave in multiphase systems. At larger thicknesses, the transmission is mainly determined by the frequency-dependent attenuation of THz waves in the ferroelectric and the formation of a standing polarization/THz wave. Our results advance the understanding of the interaction between THz wave and ferroelectrics and suggest the potential of exploiting ferroelectrics to achieve low-heat-dissipation, nonvolatile voltage modulation of THz transmission for high-data-rate wireless communication.

[*]E-mail: jhu238@wisc.edu

## I. Introduction

The increasing demand for higher-data-rate wireless communication has been driving the efforts of exploiting the abundant frequency resources in the terahertz (THz) band (frequency: 0.1-10 THz) [1,2]. To realize THz communication, the ability to modulate both the amplitude and phase of a propagating THz wave via a gate voltage is crucial. Conventional approach involves the use of semiconductors, where the application of a voltage allows tuning the local carrier concentration and hence the local conductivity and permittivity. As a result, both the amplitude [3–8] and phase [9–11] modulation can be achieved. An alternative approach is to use ferroelectric materials, which contain electric-field switchable spontaneous polarization **P**. In archetypal ferroelectrics such as $BaTiO_3$ and $PbTiO_3$, the appearance of **P** is attributed to the condensation of a soft mode phonon below the Curie temperature. In these materials, the resonance frequency of **P** (soft mode phonon) is in the THz regime [12] and can be tuned by electric field, temperature, strain, etc [13]. Therefore, **P** can resonantly interact with the incident THz wave, leading to a large polarization current $\partial \mathbf{P}/\partial t$ that in turn produces strong THz radiation. The radiated THz wave will then be superimposed with the incident THz wave, leading to an amplitude and phase modulation. Compared to semiconductors, the ferroelectrics-based approach can enable a lower-heat-dissipation and nonvolatile voltage modulation, which are both critical to on-chip thermal management.

The transmission of THz wave in ferroelectrics is critically determined by the dynamical response of **P** to a THz electric field. Experimentally, time-domain THz transmission spectroscopy has been used to measure the frequency-dependent complex dielectric permittivity of ferroelectrics in the THz band [14–21]. Contribution of different lattice modes to the permittivity can be analyzed by a classical damped oscillator dispersion model, through which key parameters such as resonant frequencies and damping coefficients can be evaluated. More recently, THz pump ultrafast x-ray diffraction has been used to directly probe the lattice/polarization dynamics inside a ferroelectric material [22–24]. Computationally, both the molecular dynamics [25–28] and the mesoscale dynamical phase-field simulations [13,24] have been used to understand and predict the THz-field-driven polarization dynamics in ferroelectrics. By utilizing a simplified Landau-Devonshire (LD) type model in terms of the polarization and the coupled equations of motion for the polarization and electromagnetic (EM) wave, analytical and numerical solutions of the reflection coefficient of a freestanding ferroelectric $BaTiO_3$ slab were obtained [29–31]. However, the LD model in these works [29–31] treats the $BaTiO_3$ as a uniaxial ferroelectric which may fail to describe the polymorphic ferroelectric phase transition. More importantly, the strain-polarization coupling terms are not included in the thermodynamic potential, which can lead to significant underestimation of the resonance frequency as will be shown below. Moreover, THz wave transmission across the $BaTiO_3$ slab, which depends on both wave reflection and absorption, was not discussed in [29–31]. Furthermore, influence of ferroelectric slab thickness on the behavior of polarization excitation and the resulting reflection, absorption, and transmission behaviors has remained virtually unexplored. In this article, we use both analytical model and direct numerical simulations to predict frequency-dependent THz wave transmission in a freestanding ferroelectric $BaTiO_3$ slab varying from a few nanometers to tens of micrometers. Our analytical model reveals the important role of strain-polarization coupling in determining the resonance frequencies of the polarization vector **P** and the significant effect of radiation-field-induced damping on the dynamics of **P**, and therefore the THz wave transmission. Spatiotemporal evolution of the coupled **P** and THz wave in the ferroelectric is obtained using dynamical phase-field simulations.

Specifically, we consider a small-amplitude incident continuous THz wave to drive the polarization oscillation within the harmonic regime. Based on the Landau-Ginzburg-Devonshire (LGD) thermodynamic potential of ferroelectrics (which is a polynomial of the vector **P** and strain/stress tensor constructed based on the crystal symmetry and incorporates strain-polarization coupling [32]) and the coupled equations of motion for the **P** and EM wave, we develop an analytical model for predicting the frequency- and thickness-dependent THz dielectric susceptibility and THz wave transmission in ferroelectrics in the regime of steady-state harmonic excitation at the thin slab limit (i.e., the slab thickness is much smaller than the THz wavelength in the slab). When the ferroelectric slab is sufficiently thin, the analytical predictions agree well with numerical simulation results obtained from a dynamical phase-field model that incorporates fully coupled dynamics of polarization, strain, and EM waves [13]. At relatively large thicknesses, it is found that the formation of standing polarization wave can lead to oscillatory thickness dependences on the THz wave transmission coefficient. Our results advance our understanding of the THz wave-matter interaction in ferroelectrics. The dynamical phase-field model is also applicable to systems with inhomogeneous polarization configurations and can therefore be used to guide the design of ferroelectrics-based active THz wave modulators for high-data-rate wireless communication applications.

## II. Analytical model of THz dielectric and transmission spectra in single-domain ferroelectrics

Let us consider the normal incidence of a single-frequency, continuous sinusoidal THz plane wave transmitting through a finite-thickness ferroelectric slab along its *z* axis, as shown in Fig. 1(a). At the initial equilibrium state, the ferroelectric slab has a spatially uniform spontaneous polarization along $+x$. The incident THz wave in the air can be written as $E_i^{\rm inc}(z,t) = E_i^{0,\rm inc}{\rm e}^{{\rm i}(k_a z-\omega t)}, i = x,y,z$, where $E_i^{0,\rm inc}$ is the amplitude of the electric-field component; $k_a$ and $\omega$ are the angular wavenumber and angular frequency in the air, respectively. The transmission coefficient *T* (reflection coefficient *R*) is calculated as the intensity of the transmitted THz wave at the surface plane $A_2$ (reflected THz wave at the surface plane $A_1$) divided by the intensity of the incident THz wave. Mathematically,

$$T = \frac{\frac{1}{2}c\kappa_0\sum_i\left(E_i^{0,{\rm A}_2}\right)^2}{\frac{1}{2}c\kappa_0\sum_i\left(E_i^{0,\rm inc}\right)^2} = \frac{\sum_i\left(E_i^{0,{\rm A}_2}\right)^2}{\sum_i\left(E_i^{0,\rm inc}\right)^2};\ R = \frac{\frac{1}{2}c\kappa_0\sum_i\left(E_i^{0,{\rm A}_1}\right)^2}{\frac{1}{2}c\kappa_0\sum_i\left(E_i^{0,\rm inc}\right)^2} = \frac{\sum_i\left(E_i^{0,{\rm A}_1}\right)^2}{\sum_i\left(E_i^{0,\rm inc}\right)^2},\ \ i = x,y,z, \quad (1)$$

where $c$ is the speed of light in the vacuum; $\kappa_0$ is the vacuum permittivity; $E_i^{0,{\rm A}_2}$ and $E_i^{0,{\rm A}_1}$ are the amplitudes of the electric-field component of the transmitted THz wave at the plane $A_2$ and the reflected THz wave at plane $A_1$, respectively. In the thin slab limit, it is rational to omit the reflection and refraction of the incident THz wave at the two ferroelectric/air interfaces and wave interference [33]. In this case, $E_i^{0,{\rm A}_2}$ is a superposition of the incident and radiated THz wave, yet $E_i^{0,{\rm A}_1}$ is contributed purely by the radiated THz wave. Since the amplitude of the incident THz wave $E_i^{0,\rm inc}$ is known, the key is to calculate the $E_i^{\rm rad}$, the radiated wave generated by the oscillating polarization in the ferroelectric.

The plane-wave solution of $E_i^{\rm rad}$ can be derived by analytically solving the Maxwell's equations (see Eq. (A12) in Appendix A). In the thin slab limit, both the oscillating polarization and $E_i^{\rm rad}$ are spatially uniform (in-phase) along the *z* axis, and the expression of $E_i^{\rm rad}$ is reduced to,

$$E_i^{\rm rad,A1}(t) = E_i^{\rm rad,A2}(t) = E_i^{\rm rad}(t) = -\frac{1}{2}\frac{d}{\kappa_0 c}\frac{\partial P_i(t)}{\partial t},\ \ i = x,y. \quad (2)$$

Note that the dynamical variation of the out-of-plane polarization component $\Delta P_z(z,t)$ only produces a non-radiating dynamical depolarization field $\Delta E_z^{\rm d}(z,t) = -\frac{1}{\kappa_0\kappa_{\rm b}}\Delta P_z(z,t)$ in the ferroelectric, which satisfies the continuity condition of electric displacement $\nabla\cdot D_i = \nabla\cdot(\kappa_0\kappa_{\rm b}E_i + P_i) = 0$. Here $\kappa_{\rm b}$ is the background dielectric permittivity of ferroelectrics, which is related to the electronic polarization and other non-permanent ionic polarization induced by the local electric field [34–36]. When the ferroelectric is excited by a small-amplitude THz wave with an angular frequency $\omega$, the excitation of nonlinear polarization can be neglected. As a result, the steady-state harmonic oscillation of the local polarization in the ferroelectric can be described by $P_i(t) = P_i^{\text{eq}} + \Delta P_i(t) = P_i^{\text{eq}} + \Delta P_i^0 \mathrm{e}^{\mathrm{i}(-\omega t+\varphi)}$, where $\Delta P_i^0$ and $\varphi$ are the amplitude and phase difference (with respect to the incident THz wave) of the polarization oscillation, respectively; $P_i^{\text{eq}}$ is the spontaneous polarization value at the initial equilibrium state. Since $P_i^{\text{eq}}$ can be obtained by minimizing the LGD potential of the ferroelectric, the key is to analytically calculate the frequency-dependent $\Delta P_i^0$ and $\varphi$ of the harmonic polarization oscillation.

To derive the formulae of $\Delta P_i^0$ and $\varphi$ in the regime of harmonic polarization oscillation, we first rewrite the equation of motion for polarization [32,37–39] into the following harmonic form,

$$\mu\frac{\partial^2 \Delta P_i}{\partial t^2}+\gamma_i\frac{\partial \Delta P_i}{\partial t} = E_i^{\text{inc},0}\mathrm{e}^{\text{-}\mathrm{i}\omega t} - \omega_i^2\mu\Delta P_i + E_i^{\text{rad}}, \qquad i = x, y, \tag{3a}$$

$$\mu\frac{\partial^2 \Delta P_i}{\partial t^2}+\gamma_i\frac{\partial \Delta P_i}{\partial t} = E_i^{\text{inc},0}\mathrm{e}^{\text{-}\mathrm{i}\omega t} - \omega_i^2\mu\Delta P_i + \Delta E_i^{\text{d}}, \qquad i = z \tag{3b}$$

where $\mu=\left(\kappa_0\omega_p^2\right)^{-1}$ is the mass coefficient (polarization inertia) and $\omega_p = \sqrt{\frac{1}{\kappa_0 V_0}\sum_n \frac{Q_n^2}{M_n}}$ is the plasma frequency, where $Q_n$ and $M_n$ are the charge and mass of the $n$th charged ion in a unit cell with volume $V_0$ [37]; $\gamma_i$ is the phenomenological damping coefficient that can be related to crystal viscosity of the ferroelectric [40]. The first term on the right of Eq. (3) assumes that the local electric fields from the incident THz wave in the ferroelectric slab are spatially uniform and in-phase, which is valid in the thin slab limit. The second term on the right of Eq. (3) describes the effective electric field arising from the Landau energy density and elastic energy density in the harmonic regime, where the resonant frequency of the polarization oscillation $\omega_i$ can be written as,

$$\omega_i = \sqrt{-\frac{A_i+B_i}{\mu}};\ \mathrm{A}_i = \left.\frac{\partial E_i^{\text{Landau}}}{\partial P_i}\right|_{P=P_i^{\text{eq}}};\ \mathrm{B}_i = \left.\frac{\partial E_i^{\text{Elas}}}{\partial P_i}\right|_{P=P_i^{\text{eq}}}, i = x, y \tag{4}$$

Since the effective electric field $E_i^{\text{Landau}} = -\left.\frac{\partial f^{\text{Landau}}}{\partial P_i}\right|_{P=P_i^{\text{eq}}}$ and $E_i^{\text{Elas}} = -\left.\frac{\partial f^{\text{Elas}}}{\partial P_i}\right|_{P=P_i^{\text{eq}}}$, the coefficients $\mathrm{A}_i$ and $\mathrm{B}_i$ are essentially the local curvature of the Landau and elastic energy density at their respective minimum point. Detailed expressions of $\mathrm{A}_i$ and $\mathrm{B}_i$ are provided in Appendix B, from which it can be seen that the coefficient $\mathrm{B}_i$ is nonzero even in a stress-free ferroelectric slab. Minimizing the Landau free energy $f^{\text{Landau}}$ with respect to the **P** allows determining the $\mathbf{P}^{\text{eq}}$ of a stress-free $BaTiO_3$ crystal, with ($P_x^{\text{eq}}$, $P_y^{\text{eq}}$, $P_z^{\text{eq}}$)=(0.26 C/m$^2$, 0, 0) at room temperature (25°C) if assuming the **P** is along the +$x$ axis (as in Fig. 1a). For this initial polarization state, the analytically calculated resonant frequency $\omega_x$ and $\omega_y$ are 2π×4.1084 THz and 2π×1.0531 THz, respectively. If excluding the $\mathrm{B}_i$, $\omega_x$ and $\omega_y$ are 2π×3.3587 THz and 2π×0.7672 THz, respectively. This large discrepancy highlights the importance of incorporating strain-polarization coupling in the calculation of resonance frequencies, despite that such piezoelectric/electrostrictive contribution has been omitted in previous theoretical works [29–31,37].

Plugging in the expressions of $E_i^{\text{rad}}$ and $\Delta E_i^{\text{d}}$ into Eq. (3a) and (3b), respectively, one has,

$$\mu\frac{\partial^2 \Delta P_i}{\partial t^2}+\gamma_i^{\text{eff}}\frac{\partial \Delta P_i}{\partial t} = -\omega_i^2\mu\Delta P_i + E_i^{\text{inc},0}\mathrm{e}^{\text{-}\mathrm{i}\omega t} \qquad i = x, y, \tag{5a}$$

$$\mu\frac{\partial^2 \Delta P_i}{\partial t^2}+\gamma_i\frac{\partial \Delta P_i}{\partial t} = -\omega_{\mathrm{eff},i}^2\mu\Delta P_i + E_i^{\mathrm{inc},0}\mathrm{e}^{-\mathbf{i}\omega t}, \qquad i = z \tag{5b}$$

where the effective damping coefficient $\gamma_i^{\mathrm{eff}} = \gamma_i + \frac{1}{2}\frac{d}{\kappa_0 c}$ and the effective resonance frequency $\omega_{\mathrm{eff},i}^2 = \omega_i^2 + \frac{1}{\mu\kappa_0\kappa_b}$. Equation (5a) indicates that $E_i^{\mathrm{rad}}$ acts as an additional source for the damping term that governs the relaxation of polarization dynamics. Equation (5b) shows that $\Delta E_i^{\mathrm{d}}$ increases the resonance frequency of the out-of-plane polarization $P_z$. Substituting the steady-state solution $\Delta P_i(t) = \Delta P_i^0\mathrm{e}^{\mathbf{i}(-\omega t+\varphi)}$ into Eq. (5) yields the following relation,

$$\Delta P_i(\omega) = \kappa_0\chi_{ii}(\omega)E_i^{\mathrm{inc},0}\mathrm{e}^{-\mathbf{i}\omega t}, \qquad \chi_{ii}(\omega) = \frac{1}{\kappa_0\left[\mu\left(\omega_i^2-\omega^2\right)-\gamma_i^{\mathrm{eff}}\mathbf{i}\omega\right]}, \qquad i = x, y, \tag{6a}$$

$$\Delta P_i(\omega) = \kappa_0\chi_{ii}(\omega)E_i^{\mathrm{inc},0}\mathrm{e}^{-\mathbf{i}\omega t}, \qquad \chi_{ii}(\omega) = \frac{1}{\kappa_0\left[\mu\left(\omega_{eff,i}^2-\omega^2\right)-\boldsymbol{i}\omega\gamma_i\right]}, \qquad i = z \tag{6b}$$

where $\chi_{ii}(\omega)$ is the diagonal component of the frequency-dependent dielectric susceptibility. We note that the expression of $\chi_{ii}(\omega)$ in Eq. (6a) is equivalent to those provided in previous theoretical works (e.g., [31,37]), except that Eq. (6a) explicitly indicates that the damping parameter of polarization oscillation $\gamma_i^{\mathrm{eff}}$ includes not only the intrinsic damping but the radiation-electric-field-induced damping. As will be shown below, it is the $\gamma_i^{\mathrm{eff}}$ (which is also thickness dependent in the thin slab limit) that determines the linewidth of the THz wave transmission spectrum.

The non-diagonal components of the dielectric susceptibility $\chi_{ij}$ $(i \neq j)$ would be zero under the assumption of harmonic oscillation for the polarization. The total dielectric permittivity $\kappa_{ii}(\omega) = \kappa_{ii}^{\mathrm{Re}}(\omega) + \mathbf{i}\kappa_{ii}^{\mathrm{Im}}(\omega)$, where the real component of the dielectric permittivity is $\kappa_{ii}^{\mathrm{Re}}(\omega) = \kappa_{\mathrm{b}} + \chi_{ii}^{\mathrm{Re}}(\omega)$ while the imaginary dielectric permittivity is $\kappa_{ii}^{\mathrm{Im}}(\omega) = \chi_{ii}^{\mathrm{Im}}(\omega)$. Under THz or lower frequency excitation where the (ferroelectric) polarization **P** can follow the applied electric field **E**, the anisotropy in the response of the dielectric displacement **D** to **E** is mainly determined by **P**. In this regard, $\kappa_{\mathrm{b}}$ is typically considered to be isotropic and frequency independent. Hereafter, we restrict our discussion to the case where the $P_i^{\mathrm{eq}}$ is completely in the *xy* plane ($P_z^{\mathrm{eq}}$=0) for two reasons. First, stabilizing an out-of-plane polarization $P_z$ in a thin ferroelectric slab typically requires the use of metallic electrodes to screen the polarization charges at the surfaces, in which case the incident THz wave would be largely reflected by the electrodes. Second, as mentioned above, a time-varying $P_z$ does not produce EM radiation propagating along the out-of-plane direction under the plane-wave assumption. From Eq. (6a), we can also derive,

$$\Delta P_i^0(\omega) = \mathrm{sgn}\left(\frac{\sin\varphi}{\gamma_i^{\mathrm{eff}}}\right)E_i^0\sqrt{\frac{1}{\left(\omega_i^2\mu-\mu\omega^2\right)^2+\left(\gamma_i^{\mathrm{eff}}\omega\right)^2}}, i = x, y \tag{7a}$$

$$\varphi(\omega) = \mathrm{atan}\left[\frac{\gamma_i^{\mathrm{eff}}\omega}{\mu\left(\omega_i^2-\omega^2\right)}\right], \text{when } \omega < \omega_i;\ \varphi(\omega) = \mathrm{atan}\left[\frac{\gamma_i^{\mathrm{eff}}\omega}{\mu\left(\omega_i^2-\omega^2\right)}\right] + \pi, \text{when } \omega > \omega_i \tag{7b}$$

Now consider the normal incidence of a continuous THz wave with an electric-field component along the *x* axis ($E_x^{\mathrm{inc}} \neq 0$, $E_y^{\mathrm{inc}} = E_z^{\mathrm{inc}} = 0$) as an example. In this case, only the $P_x$ will oscillate, hence the radiation electric field $E_i^{\mathrm{rad}}$ only has the *x*-component. Thus, the electric field at the plane $A_2$ can be expressed as,

$$E_x^{\mathrm{A}_2} = E_x^{0,\mathrm{inc}} \mathrm{e}^{-\mathbf{i}\omega t} + \frac{1}{2}\frac{d}{\kappa_0 c}\mathbf{i}\omega\Delta P_x^0 \mathrm{e}^{\mathbf{i}(-\omega t+\varphi)}. \tag{8}$$

The temporal waveform of $E_x^{\mathrm{A}_2}(t)$ is the real component of the complex function in Eq. (8), i.e.,

$$E_x^{\mathrm{A}_2}(t) = E_x^{0,\mathrm{inc}} \cos(\omega t) + \frac{1}{2}\frac{d}{\kappa_0 c}\omega\Delta P_x^0 \sin(\omega t - \varphi). \tag{9}$$

Since both $E_x^{0,\mathrm{inc}}$ and $\omega$ are the inputs, and the analytical solutions of $\Delta P_x^0$ and $\varphi$ are provided in Eqs. (7a-b), one can analytically predict the entire temporal waveform and therefore the amplitude $E_x^{0,\mathrm{A}_2}$. Accordingly, the transmission coefficient can be calculated by $T = \left(E_x^{0,\mathrm{A}_2}/E_x^{0,\mathrm{inc}}\right)^2$. The reflection coefficient can be calculated as $R = \left(E_x^{0,\mathrm{A}_1}/E_x^{0,\mathrm{inc}}\right)^2$. The reflected electric field at the plane $\mathrm{A}_1$ is the same as the $E_i^{\mathrm{rad}}$ in Eq. (2) (see Eq. (A12) in Appendix A for the exact solution of $E_i^{\mathrm{rad}}$), and thus the amplitude $E_x^{0,\mathrm{A}_1}$ can be analytically calculated. We note that $T + R < 1$ because the damping of polarization wave (via the $\gamma_i$ in Eq. (3)) leads to THz wave absorption.

**III. Dynamical phase-field simulations**

In this section we present our dynamical phase-field model which allow numerical simulation of THz transmission, reflection, and absorption in ferroelectrics with spatially inhomogeneous polarization. The propagation of THz wave is coupled to both polarization dynamics and strain dynamics in a ferroelectric system. Therefore, the simulation of THz wave transmission requires solving the coupled equations of motion for EM waves, polarization **P**, and the mechanical displacement **u**. Rearranging the Maxwell's equations, one has,

$$\frac{\partial \mathbf{E}}{\partial t} = \frac{1}{\kappa_0 \kappa_\mathrm{b}}\left(\nabla \times \mathbf{H} - \mathbf{J}^\mathrm{f} - \frac{\partial \mathbf{P}}{\partial t}\right), \tag{10}$$

$$\frac{\partial \mathbf{H}}{\partial t} = -\frac{1}{\mu_0}(\nabla \times \mathbf{E}). \tag{11}$$

It is noteworthy that Eqs. (10-11) are solved for the entire system shows in Fig. 1(a), which includes both the ferroelectric phase and the air phase. As indicated by Eq. (10), the electric-field component of the EM wave (**E**) is related to both the free electric current $\mathbf{J}^\mathrm{f}$ and the polarization current $\partial\mathbf{P}/\partial t$. We set the background dielectric permittivity $\kappa_\mathrm{b}$=5 [41] in the ferroelectric $BaTiO_3$, and $\kappa_\mathrm{b}$=1 in the air. The evolution of the polarization **P** is governed by,

$$\mu\frac{\partial^2 P_i}{\partial t^2} + \gamma_i \frac{\partial P_i}{\partial t} = E_i^{\mathrm{eff}}, \tag{12}$$

where the total effective electric field $E_i^{\mathrm{eff}} = E_i^{\mathrm{Landau}} + E_i^{\mathrm{Elas}} + E_i^{\mathrm{Grad}} + E_i^{\mathrm{d}} + E_i^{\mathrm{ext}}$ include effective electric fields arising ($E_i^{\mathrm{Landau}}$ and $E_i^{\mathrm{Elas}}$) from the Landau free energy and elastic energy densities, the field from the gradient energy density $E_i^{\mathrm{Grad}}$, the depolarization field $E_i^{\mathrm{d}}$, and the externally applied electric field $E_i^{\mathrm{ext}}$. The expression of $E_i^{\mathrm{Landau}}$ (available in Appendix B) is a seventh-order polynomial of the polarization $P_i$, and thus Eq. (12) would describe an anharmonic polarization oscillation if the amplitude of $\Delta P_i$ is large enough. The expression of $E_i^{\mathrm{Elas}} = -\frac{\partial f^{\mathrm{Elas}}}{\partial P_i}$ is also provided in Appendix B. The elastic energy density $f^{\mathrm{Elas}} = \frac{1}{2}c_{ijkl}\left(\varepsilon_{kl} - \varepsilon_{kl}^0\right)\left(\varepsilon_{ij} - \varepsilon_{ij}^0\right)$, where $c_{ijkl}$ is the elastic stiffness tensor of the ferroelectric.

$\boldsymbol{\varepsilon}^0$ is the stress-free strain, calculated as $\varepsilon_{ii}^0 = Q_{11}P_i^2 + Q_{12}\left(P_j^2 + P_k^2\right)$ and $\varepsilon_{ij}^0 = Q_{44}P_iP_j$, with $i = x, y, z$, and $j \neq i$ [42]. $Q_{11}$, $Q_{12}$, and $Q_{44}$ are the electrostrictive coefficients of the ferroelectric. The total strain $\boldsymbol{\varepsilon}$ can be written as $\boldsymbol{\varepsilon} = \boldsymbol{\varepsilon}^{\mathrm{eq}} + \Delta\boldsymbol{\varepsilon}$. $\boldsymbol{\varepsilon}^{\mathrm{eq}}$ is the total strain at the initial equilibrium state, which describes the macroscopic shape change of the ferroelectric and can be obtained by solving the mechanical equilibrium equation $\nabla\cdot\boldsymbol{\sigma}^{\mathrm{eq}} = 0$. Here $\boldsymbol{\sigma}^{\mathrm{eq}}$ is the stress distribution at the initial equilibrium state. If the polarization is spatially uniform in a stress-free ferroelectric slab, $\boldsymbol{\varepsilon}^{\mathrm{eq}} = \boldsymbol{\varepsilon}^{0,\mathrm{eq}}$. The dynamical strain $\Delta\boldsymbol{\varepsilon}$, which originates from the oscillating polarization, can be obtained by numerically solving the elastodynamic equation,

$$\rho\frac{\partial^2\Delta\mathbf{u}}{\partial t} = \nabla\cdot\left(\Delta\boldsymbol{\sigma} + \beta\frac{\partial\Delta\boldsymbol{\sigma}}{\partial t}\right). \quad (13)$$

Here $\rho$ is the mass density and $\beta$ is the stiffness damping coefficient. $\Delta\mathbf{u} = \mathbf{u} - \mathbf{u}^{\mathrm{eq}}$ is the dynamical displacement and $\Delta\boldsymbol{\sigma} = \boldsymbol{\sigma} - \boldsymbol{\sigma}^{\mathrm{eq}}$ is the dynamical stress. One can also write $\Delta\sigma_{ij} = c_{ijkl}\left(\Delta\varepsilon_{kl} - \Delta\varepsilon_{kl}^0\right)$, with $\Delta\varepsilon_{kl}^0 = \varepsilon_{kl}^0 - \varepsilon_{kl}^{0,\mathrm{eq}}$ and $\Delta\varepsilon_{kl} = \frac{1}{2}\left(\frac{\partial\Delta u_k}{\partial l} + \frac{\partial\Delta u_l}{\partial k}\right)$. $E_i^{\mathrm{Grad}} = -\frac{\delta f^{\mathrm{grad}}}{\delta P_i} = G\nabla^2 P_i$, where $G$ is the isotropic gradient energy coefficient. $E_i^{\mathrm{ext}}$ is the same as the electric-field component of the propagating EM wave in the system, which is obtained by averaging the simulated $E_i$ staggered on the edges of the Yee lattice. $E_i^{\mathrm{d}}$ can be expressed as $E_i^{\mathrm{d}}(t) = E_i^{\mathrm{d,eq}} + \Delta E_i^{\mathrm{d}}(t)$. The depolarization field at the initial equilibrium state ($t$=0), $E_i^{\mathrm{d,eq}}$, can be obtained by solving the electrostatic equilibrium equation $\nabla\cdot\left(\kappa_0\kappa_{\mathrm{b}}E_i^{\mathrm{d,eq}} + P_i^{\mathrm{eq}}\right) = 0$. For a 1D system which has periodic boundary condition within the $xy$ plane, one has $\left(E_x^{\mathrm{d,eq}}, E_y^{\mathrm{d,eq}}, E_z^{\mathrm{d,eq}}\right) = \left(0, 0, -\frac{P_z^{\mathrm{eq}}}{\kappa_0\kappa_{\mathrm{b}}}\right)$. The dynamical depolarization field $\Delta E_i^{\mathrm{d}}$ does not need to be calculated separately. Rather, the $E_i(t)$ obtained by numerically solving Eqs. (10) and (11) via the finite-difference time-domain (FDTD) solver on a Yee lattice can automatically satisfy the condition of $\nabla\cdot D_i = 0$. The stress-free mechanical boundary condition at the top and bottom surfaces of the ferroelectric layer is automatically satisfied after setting the elastic stiffness coefficient $c_{ijkl} \approx 0$ for the air phase.

The entire system is discretized into one-dimensional (1D) cells of $1\times1\times\Delta zN$ with a cell size $\Delta z$= 1 nm. The bottom 40 cells (40 nm) and the top 20 cells (20 nm) are the air phase, where $\rho \approx 0$, $\beta \approx 0$, $c_{ijkl} \approx 0$. The remaining cells are designated the ferroelectric $BaTiO_3$. The free electric current source is injected at the coordinate $z$=-20. The boundary condition $\partial P_i/\partial z = 0$ ($i$=x,y,z) is used when solving Eq. (12), which assumes that the magnitude of polarization at the surface is the same as bulk value (i.e., the extrapolation length [43,44] is infinity). This Neumann type boundary condition would lead to the reflection of polarization wave at the two surfaces of the ferroelectric slab. Conventional Yee lattice is used to discretize the EM wave and FDTD is used to solve Eqs. (10-11), during which the absorbing boundary condition, $\frac{\partial\mathbf{E}}{\partial z} = -\frac{1}{c}\frac{\partial\mathbf{E}}{\partial t}$, is applied to the topmost and bottommost cell of the entire computational system. Due to the use of the staggered grids for $\boldsymbol{E}$ and $\boldsymbol{H}$ in the Yee lattice (see Fig. 1(a), right panel), the actual thickness of the $BaTiO_3$ layer considered in numerical simulations would be ($N$-60+1)$\Delta z$, where the additional grid results from the two half grids at the interfaces. Central finite difference is used for calculating spatial derivatives. All dynamical equations are solved simultaneously using the classical Runge-Kutta method for time-marching with a real-time step $\Delta t = 2\times10^{-18}$ s. The high numerical accuracy of our FDTD solver is demonstrated by simulating the EM wave transmission across a freestanding dielectric layer with spontaneous polarization **P**=0 and comparing the simulation results with analytical solution derived from the ♯SHAARP package [45] (see Appendix C). To evaluate the transmission coefficient from the numerical simulations, we first obtain the $E_i^{0,\mathrm{inc}}$ by a control simulation with **P**=0 and $\kappa_{\mathrm{b}} = 1$, and then record the $E_i^{0,A_2}$ at the plane $A_2$ after the polarization oscillation reaches the steady-state regime. The transmission coefficient can then be evaluated via Eq. (1). The radiation electric field $E_i^{\mathrm{rad}}(z,t)$ can be extracted by subtracting the $E_i^{\mathrm{inc}}(z,t)$ from the total $E_i(z,t)$.

The material parameters of $BaTiO_3$ single crystal used in the analytical calculations and dynamical phase-field simulations are summarized as follows. The elastic stiffness coefficients [46] $c_{11}$ = 178 GPa, $c_{12}$ = 96.4 GPa and $c_{44}$ = 122 GPa; the mass density $\rho$ = 6020 kg $m^{-3}$; the mass coefficient $\mu = 1.35\times10^{-18}$ J m $s^2$ $C^{-2}$ [37]; the coefficients in the Landau free energy density $f^{\mathrm{Landau}}$ [47]: $\alpha_1(T) = 4.124\times10^5\times(T\text{-}115)$ $C^{-2}$ $m^2$ N, $\alpha_{11} = -2.097\times10^8$ $C^{-4}$ $m^6$ N, $\alpha_{12} = 7.974\times10^8$ $C^{-4}$ $m^6$ N, $\alpha_{111} = 1.294\times10^9$ $C^{-6}$ $m^{10}$ N, $\alpha_{112} = -1.95\times10^9$ $C^{-6}$ $m^{10}$ N and $\alpha_{123} = -2.5\times10^9$ $C^{-6}$ $m^{10}$ N, $\alpha_{1111} = 3.863\times10^{10}$ $C^{-8}$ $m^{14}$ N, $\alpha_{1112} = 2.529\times10^{10}$ $C^{-8}$ $m^{14}$ N, $\alpha_{1122} = 1.637\times10^{10}$ $C^{-8}$ $m^{14}$ N, $\alpha_{1123} = 1.367\times10^{10}$ $C^{-8}$ $m^{14}$ N and temperature $T$ is in °C; the isotropic gradient energy coefficient [48] $G_{11} = 5.1\times10^{-10}$ J $m^3$ $C^{-2}$; the electrostrictive coefficients [47] $Q_{11}$ = 0.1 $C^{-2}$ $m^4$, $Q_{12}$ = -0.034 $C^{-2}$ $m^4$, and $Q_{44}$ = 0.029 $C^{-2}$ $m^4$.

**IV. Results and Discussion**

To generate a continuous incident THz wave with an electric-field component $E_i$, we inject a sinusoidal source current in the form of $J_i^{\mathrm{f}} = J_i^0\sin(\omega t)$, $i$=$x$,$y$,$z$, where $J_i^0$ is the amplitude and $\omega$ is the angular frequency. Since the focus of this work is linear THz wave transmission, $J_i^0$ needs to be small enough such that the polarization oscillation is within the harmonic regime. Figure 1(b) shows the sum of Landau and elastic energy density ($f^{\mathrm{s}}$=$f^{\mathrm{Landau}}$+$f^{\mathrm{Elast}}$) as a function of $P_x$ for $P_y$=$P_z$=0 (top) and of $P_y$ for $P_x$=$P_x^{\mathrm{eq}}$, $P_z$=0 (bottom) as well as the harmonic fitting via $f^{\mathrm{s}}(P_i) = \frac{1}{2}(A_i + B_i)\left(P_i - P_i^{\mathrm{eq}}\right)^2$, $i$=$x$,$y$, where $A_i$ and $B_i$ are the analytically calculated local curvature of the $f^{\mathrm{Landau}}$ and $f^{\mathrm{Elas}}$ at the initial equilibrium state ($P_x^{\mathrm{eq}}$, $P_y^{\mathrm{eq}}$, $P_z^{\mathrm{eq}}$)=(0.26 C/$m^2$, 0, 0) at 25°C. The fitting suggests a harmonic regime of variation of ±4.89 mC/$m^2$ for $P_x$ and ±9.6 mC/$m^2$ for $P_y$. In dynamical phase-field simulations, we set $J_i^0$= $1\times10^{11}$ A/$m^2$. The resulting electric-field amplitude is $E_i^{0,\mathrm{inc}}$=18836 V/m ($i$=$x$, $y$), which ensures that $\Delta P_i^0$ is within the harmonic regime even under the resonant excitation condition ($\omega$=$\omega_i$). By letting the initial polarization state deviates slightly from the equilibrium state of, the subsequent damped polarization oscillation processes with and without the $E_i^{\mathrm{Elas}}$ are simulated and shown in Fig. 1(c) and 1(d), where the phenomenological damping coefficient $\gamma_x$=$\gamma_y$=$\gamma$=$2\times10^{-7}$ Ω·m and the thickness of (100) $BaTiO_3$ is 10 nm. The values of resonance (soft mode) frequencies extracted from their frequency spectra, as shown on the right of the time-domain plot, agree remarkably well with the analytical predictions via Eq. (4).

Figure 2(a) shows the simulated steady-state evolution of the $E_x^{\mathrm{inc}}$ and the resultant $\Delta P_x$ at the topmost cell of a 10-nm-thick (100) $BaTiO_3$ slab at three different angular frequencies $\omega$ of the incident THz wave, from which the peak amplitude $\Delta P_x^0$ and phase $\varphi$ (with respect to the incident wave) can be extracted. During evolution, both the polarization and the THz wave are spatially uniform at such a small thickness (10 nm), thus the simulation results should be consistent with analytical predictions via the formulae derived in Sect. II. Figure 2(b) shows the frequency-dependent $|\Delta P_x^0|$ and $\varphi$ under three different values of damping coefficient $\gamma$. $|\Delta P_x^0|$ is smaller under larger $\gamma$ due to larger energy dissipation in the lattice of $BaTiO_3$, and the corresponding $\varphi$ is shown in the inset. The simulated $|\Delta P_x^0|(\omega)$ and $\varphi(\omega)$ both agree well with those calculated analytically via Eqs. (7a-b). The $|\Delta P_x^0|$ shows a resonant enhancement at $\omega$=$\omega_x$=2π×4.1084 THz when $\gamma$ is relatively small ($2\times10^{-7}$ Ω·m and $2\times10^{-5}$ Ω·m). However, the $|\Delta P_x^0|$ decreases monotonically with increasing $\omega$ when $\gamma$ is relatively large ($6\times10^{-5}$ Ω·m), because the effective damping coefficient $\gamma^{\mathrm{eff}} = \gamma + \frac{1}{2}\frac{d}{\kappa_0 c}$ in this case (~$6.19\times10^{-5}$ Ω·m) is near the threshold (=$2\omega_x\mu$=$6.97\times10^{-5}$ Ω·m) for overdamping (as discussed in [13]). We also applied the source current $J_i^{\mathrm{f}}$ along the $y$ axis to produce an incident THz electric field of the same peak amplitude along $y$. The initial equilibrium polarization is still along $+x$. In this case, the numerically simulated $\Delta P_i$ is mainly along the $y$ axis, although a small $\Delta P_x$ (on the order of $10^{-7}$ C/m$^2$) is induced by the nonzero $E_x^{\mathrm{Landau}}$. Figure 2(b) shows the $|\Delta P_y^0|(\omega)$ and $\varphi(\omega)$ obtained from both analytical calculations and dynamical phase-field simulations, which are well consistent with each other. Likewise, the $|\Delta P_y^0|$ shows a resonant enhancement at $\omega=\omega_y=2\pi\times1.0531$ THz when $\gamma$ is relatively small ($2\times10^{-7}$ Ω m) yet deceases monotonically with increasing $\omega$ at larger $\gamma$. Since the threshold $\gamma^{\mathrm{eff}}$ for overdamping is ~$1.79\times10^{-5}$ Ω·m ($=2\omega_y\mu$), the polarization oscillation becomes overdamped for the cases of $\gamma=2\times10^{-5}$ Ω·m and $6\times10^{-5}$ Ω·m.

Knowing $|\Delta P_x^0|(\omega)$ and $\varphi(\omega)$ allows evaluating the total electric field at the A$_2$ plane $E_x^{A_2}(t)$ via Eq. (9), and thus the calculation of the transmission coefficient. Figure 3(a) and 3(b) show the frequency-dependent transmission of the incident THz wave $E_x^{\mathrm{inc}}$ and $E_y^{\mathrm{inc}}$ under different $\gamma$ values in a 10-nm-thick (100) $BaTiO_3$ film, respectively. The peak amplitude of the source current is $J_i^0=1\times10^{11}$ A/m$^2$ ($i=x,y$). The numerically simulated transmission coefficients agree remarkably well with the analytically calculated values. As shown in Figs. 3(a-b), the transmission coefficient is larger at smaller $\gamma$ value under off-resonance condition ($\omega \ll \omega_i$ or $\omega \gg \omega_i$). In this case, the transmission is dominated by the phase difference $\varphi$. Take the regime of $\omega \ll \omega_x$ as an example, a smaller $\gamma$ leads to a larger $|\Delta P_x^0|$ (Fig. 2(b)) and therefore a larger amplitude for the $E_x^{\mathrm{rad}}$. However, the phase difference $\varphi$ is smaller at smaller $\gamma$ values (Fig. 2(b) inset). Accordingly, the destructive interference, described by the second term on the right-hand side of Eq. (9), is less significant, and thus a larger transmission is achieved. For near ($\omega\sim\omega_i$) or on-resonance ($\omega=\omega_i$) condition, the transmission coefficient is smaller at smaller $\gamma$ values. In this regime, the phase difference $\varphi$ is near or equal to π/2 (see Eq. (7b)). Therefore, the transmission is dominated by the amplitude $|\Delta P_x^0|$. Specifically, the larger $|\Delta P_x^0|$ at smaller $\gamma$ value leads to a larger amplitude of $E_x^{\mathrm{rad}}$ and thus more significant destructive interference. As a result, the transmission coefficient is smaller.

Figure 4(a,e) show both the analytically calculated and numerically simulated thickness-dependent transmission coefficients of the incident THz wave $E_x^{\mathrm{inc}}$ and $E_y^{\mathrm{inc}}$ at three different frequency values, and Figures 4(b,f) show the zoom-in plot at lower thicknesses. The $\gamma$ is $2\times10^{-7}$ Ω·m, and the $J_i^0$ is $1\times10^{11}$ A/m$^2$ ($i=x,y$) in all cases. As mentioned previously, the analytical model (via Eq. (9)) assumes both the polarization and the THz wave are spatially uniform during evolution, which can only be true when the film thickness $d$ is significantly smaller than THz wavelength in the ferroelectric slab $\lambda_0 = \frac{2\pi}{\mathbf{k}^{\mathrm{Re}}}$. Here $\mathbf{k}^{\mathrm{Re}}$ is the real part of the complex wavenumber $\mathbf{k}$ of the incident THz wave in the ferroelectric slab, with $k_i = k_i^{\mathrm{Re}} + \mathbf{i}k_i^{\mathrm{Im}} = \omega\sqrt{\kappa_0\kappa_{ii}(\omega)\mu_0}$, $i=x,y$, where $\kappa_{ii}(\omega) = \kappa_{\mathrm{b}} + \chi_{ii}(\omega)$. Assuming that $\kappa_{ii}(\omega)$ in the thick slab is the same as that in the thin slab limit and that $\gamma_i^{\mathrm{eff}} \approx \gamma_i$, we analytically evaluate the frequency-dependent $k_i^{\mathrm{Re}}$ and $k_i^{\mathrm{Im}}$, as shown in Figs. 4(c,d) for $E_x^{\mathrm{inc}}$ and Figs. 4(g,h) for $E_y^{\mathrm{inc}}$. From the $k_i^{\mathrm{Re}}(\omega)$, we obtain that $\lambda_0$=7.838 μm, 0.698 μm, and 2701.53 μm for $E_x^{\mathrm{inc}}$ at 2π×2.6084 THz, 2π×4.1084 THz ($=\omega_x$), and 2π×8.1084 THz, respectively, and $\lambda_0$=10.550 μm, 1.379 μm, and 2536.878 μm for $E_y^{\mathrm{inc}}$ at 2π×0.5531 THz, 2π×1.0531 THz ($=\omega_y$), and 2π×5.0531 THz, respectively.

Since the thin slab limit ($d \ll \lambda_0$) is readily reached when $\omega \gg \omega_i$, good consistency between analytical model and numerical simulations is achieved in this frequency regime, see for example the case of 2π×8.1084 THz for $E_x^{\mathrm{inc}}$ and 2π×5.0531 THz for $E_y^{\mathrm{inc}}$, as shown in Fig. 4(a,e). At $\omega = \omega_i$, while the simulation results still agree well with analytical calculations when $d$ is small enough (e.g., $d$<200 nm for $E_x^{\rm inc}$ at 2π×4.1084 THz, Fig. 4(b)), significant deviation appears at larger thicknesses, as shown more clearly in Figs. 4(b,f). At $\omega < \omega_i$, oscillatory behaviors appear, in sharp contrast with the monotonically decreasing transmission coefficient with increasing thickness at $\omega \geq \omega_i$. Such oscillatory behaviors are attributed to the stronger attenuation of higher-frequency THz waves and the formation of standing polarization/THz waves in the lower-frequency cases, which we elaborate from the following two aspects. First, by writing the incident THz wave as $E_i^{\rm inc} = E_i^{0,\rm inc} e^{\mathbf{i}(\mathbf{k}z-\omega t)} = E_i^{0,\rm inc} e^{-k_i^{\rm Im} z} e^{-\mathbf{i}(k_i^{\rm Re} z-\omega t)}$ (*i*=*x*,*y*), one can see that the imaginary part of the wavenumber $\mathbf{k}^{\rm Im}$determines the attenuation for the incident wave. From the analytically calculated $k_i^{\rm Im}(\omega)$ shown in Figs. 4(d,h), $k_i^{\rm Im}$ peaks at the resonant frequency ($\omega$=$\omega_i$), indicating the strongest wave attenuation. Moreover, $k_i^{\rm Im}$ at above $\omega_i$ is orders-of-magnitude larger than those below $\omega_i$, indicating a stronger attenuation for higher-frequency THz wave, which is consistent with the monotonic decrease in transmission at 2π×8.1084 THz (5.0531 THz) for $E_x^{\rm inc}$ ($E_y^{\rm inc}$). Second, the relatively low $k_i^{\rm Im}$ at below $\omega_i$ suggests that the incident THz wave can propagate across the entire film thickness and interfere with the reflected THz waves. It is this interference that leads to the oscillatory transmission coefficient seen at 2π×2.6084 THz (0.5531 THz) for $E_x^{\rm inc}$ ($E_y^{\rm inc}$). More detailed analyses show that the first, second, and third peaks in the oscillatory transmission curve correspond to the formation of *n*=1, 2, 3 mode standing THz wave (i.e., $\frac{n\pi}{d} = k^{\rm Re}$), respectively (see Appendix D). This phenomenon can be explained by approximately writing the standing THz wave to be $E_i^{\rm stand} \approx Ae^{-k^{\rm Im}z}e^{-\mathbf{i}(k^{\rm Re}z-\omega t)} + Be^{-k^{\rm Im}(2d-z)}e^{-\mathbf{i}(-k^{\rm Re}z-\omega t)}$ with 0≤*z*≤*d*, from which it can be seen that $E_i^{\rm stand}$ peaks when $z = \frac{n\pi}{k^{\rm Re}}$.

## V. Conclusions

We have developed an LGD-theory based analytical model for predicting the frequency- and thickness-dependent THz wave transmission across ferroelectrics at the thin slab limit in the regime of steady-state harmonic excitation. The analytical model indicates that the polarization-current-induced radiation electric field increases the effective damping coefficient for polarization dynamics. On one hand, a larger effective damping reduces the amplitude of polarization excitation, leading to a higher THz wave transmission. On the other hand, a larger effective damping can result in a lower THz wave transmission by modulating the phase difference between the polarization oscillation and the incident electric field. Which mechanism dominates the transmission depends on whether the THz wave frequency is near or off resonance with the ferroelectric polarization. When the ferroelectric slab thickness is significantly smaller than the THz wavelength inside the ferroelectrics, the predictions from the analytical model agree well with simulation results from a dynamical phase-field model, which incorporates coupled dynamics of polarization, strain, and EM waves in multiphase systems. At large thicknesses, the transmission behavior is governed by the propagation, attenuation, and reflection of THz wave inside the ferroelectric as well as the resulting formation of standing polarization/THz waves, which is revealed by our dynamical phase-field simulations.

The strong absorption at or near the resonant frequency suggests the potential of realizing an effective electric-field control of the THz wave transmission through the modulation of the spontaneous polarization at the initial equilibrium and the resultant resonant frequency. Our analytical model can be used to analyze experimentally measured steady-state THz wave transmission spectra in single-domain ferroelectrics in the harmonic regime and thin slab limit, which allow extracting key parameters such as the mass and damping coefficients of polarization dynamics and probe the local curvature of the free energy landscape. However, since the present analytical solutions are derived for the steady-state polarization oscillation, transmission experiments based on continuous wave THz sources would be needed to validate the prediction. To understand THz transmission experiments that involve the use of single-cycle, picosecond-duration broadband THz pulse (as in time-domain THz transmission spectroscopy [14–21]), the present analytical model needs to be adapted to describe transient-state polarization dynamics. Alternatively, one can directly use our dynamical phase-field model which permits simulating THz transmission in more general cases including systems with inhomogeneous polarization configuration such as polar vortices [49] and skyrmions [50]. The dynamical phase-field model can be used to design the polarization domain structure of single-phase ferroelectrics or ferroelectric heterostructures (e.g., multilayer, superlattices) for realizing on-demand control over the propagation, transmission, and reflection of THz wave for high-data-rate wireless communication. Both the analytical theory and dynamical phase-field models can be generalized to investigate THz wave transmission for other polar materials that have spontaneous polarization and piezoelectric effect such as III-nitride semiconductors.

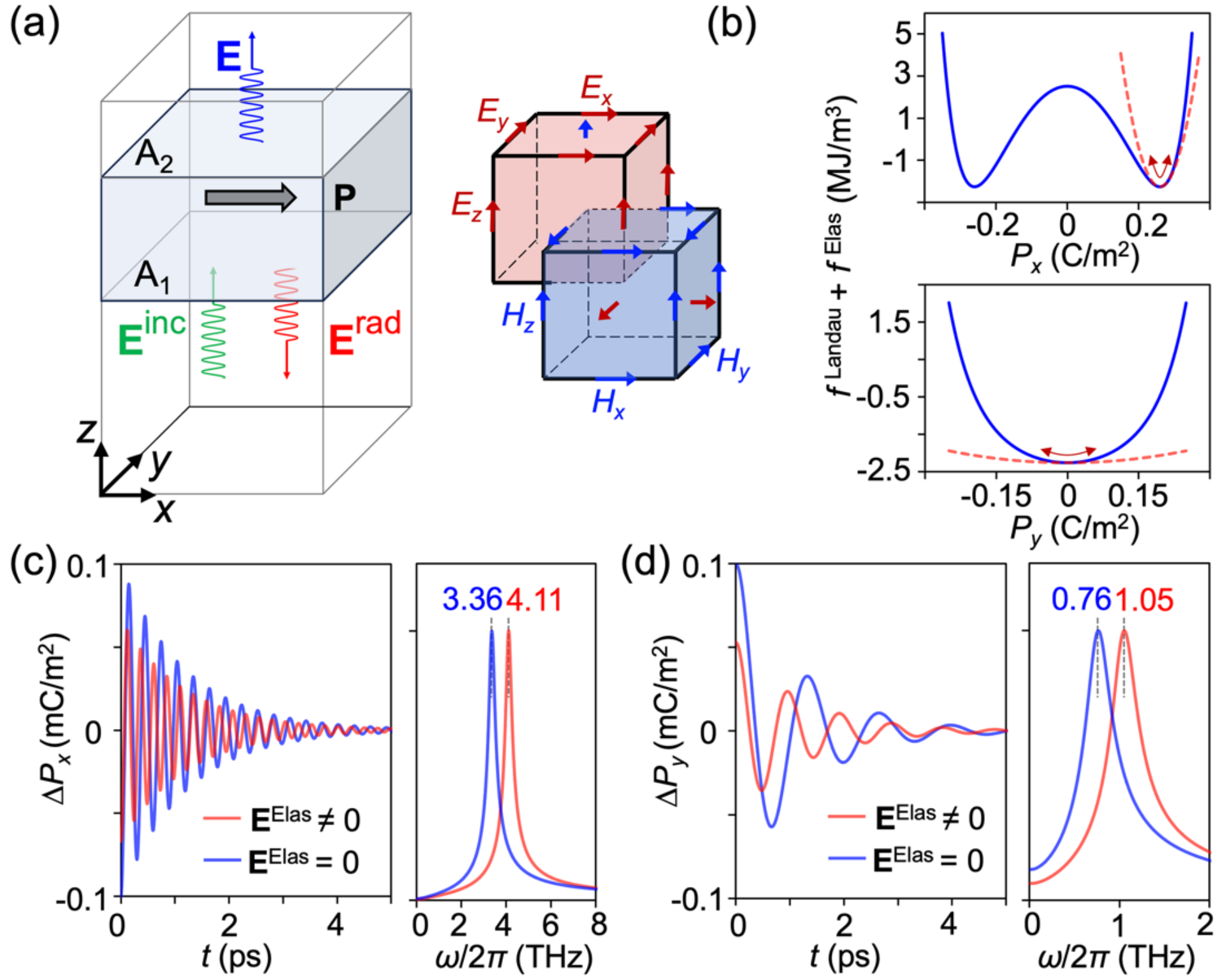

**Figure 1. (a)** Schematic of the system set-up for studying THz wave transmission across a freestanding ferroelectric layer with a single-domain spontaneous polarization **P** at the initial equilibrium state. The incident THz wave ($\mathbf{E}^{\mathrm{inc}}$) is produced by a free charge current source $\mathbf{J}^{\mathrm{f}}$. The electric-field component of the transmitted THz wave **E** is a superposition of the $\mathbf{E}^{\mathrm{inc}}$ and the polarization-current ($\partial\mathbf{P}/\partial t$) induced radiation electric field $\mathbf{E}^{\mathrm{rad}}$. The reflected THz wave is $\mathbf{E}^{\mathrm{rad}}$. Shown on the right is the schematic of the Yee lattice where staggered cells are used for discretization of the **E** and **H** component of the THz wave in our dynamical phase-field model. (**b**). Thermodynamic potential energy density as a function of $P_x$ for $P_y=P_z=0$ (top) and of $P_y$ for $P_x=P_x^{\mathrm{eq}}$, $P_z=0$ (bottom) as well as the harmonic fitting (dashed lines) in a freestanding (100) $BaTiO_3$ slab; (**c**,**d**) Temporal profile of oscillatory polarization relaxation near the equilibrium polarization state at the minima of the energy landscape shown in (**b**), with and without considering the elastic effective electric field $\mathbf{E}^{\mathrm{Elas}}$.

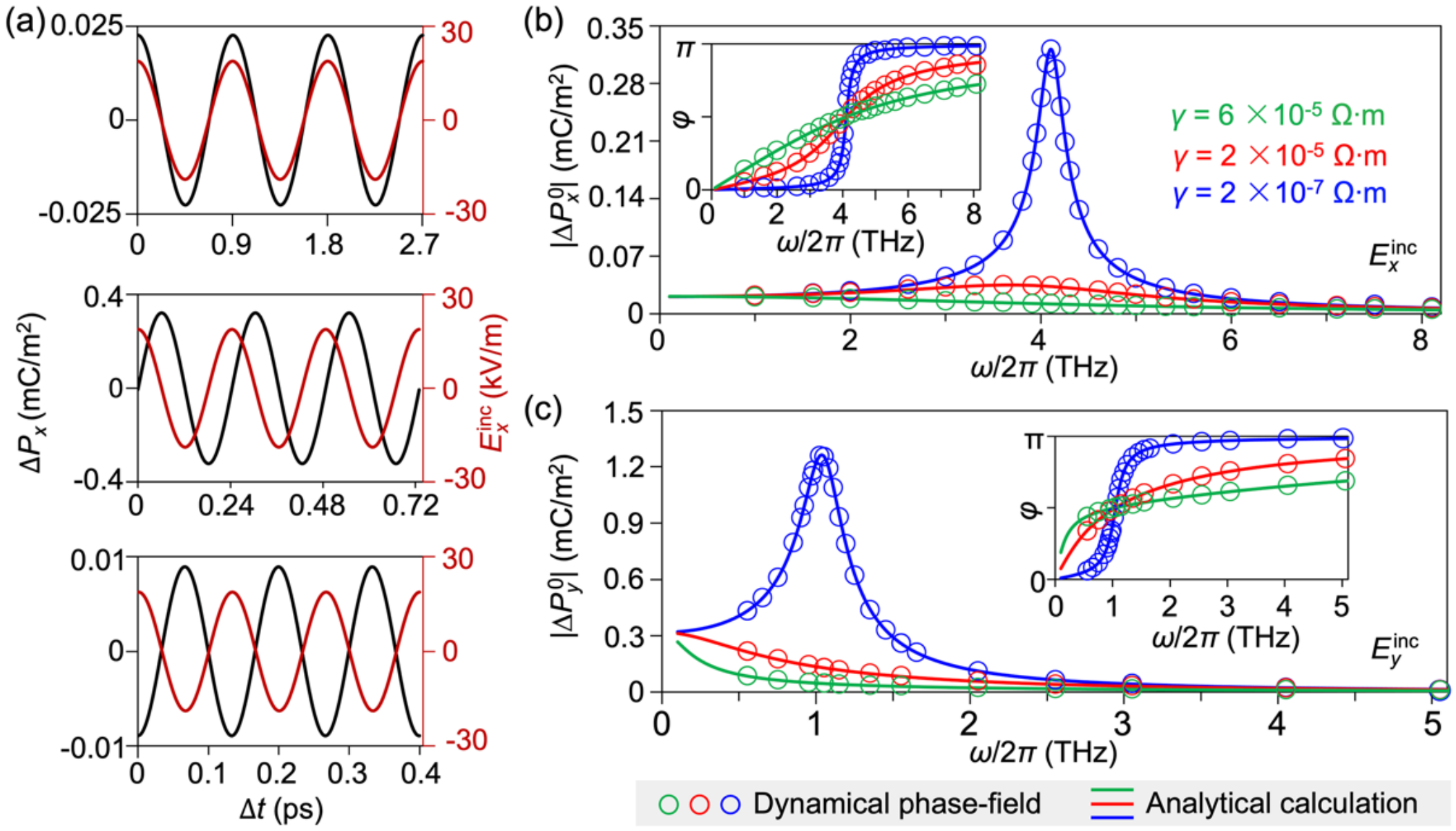

**Figure 2**. (**a**) Numerically simulated steady-state evolution of the $E_x^{\text{inc}}$ and the resultant $\Delta P_x$ at the topmost cell of a 10-nm-thick (100) $BaTiO_3$ slab. (From top to bottom) the angular frequency of $E_x^{\text{inc}}$ $\omega$=2π× 1.1084 THz, 2π× 4.1084 THz (=$\omega_x$, on-resonance), and 2π× 7.5084 THz, respectively. Amplitude $|\Delta P_i^0|$ and phase $\varphi$ (inset) of the steady-state harmonic polarization oscillation $\Delta P_i(t) = \Delta P_i^0 e^{\mathrm{i}(-\omega t+\varphi)}$ in a 10-nm-thick (100) $BaTiO_3$ slab under the excitation of $E_i^{\text{inc}}$ of different angular frequency $\omega$ with different damping coefficient $\gamma$, with the subscript *i*=*x* for (**b**) and *i*=*y* for (**c**).

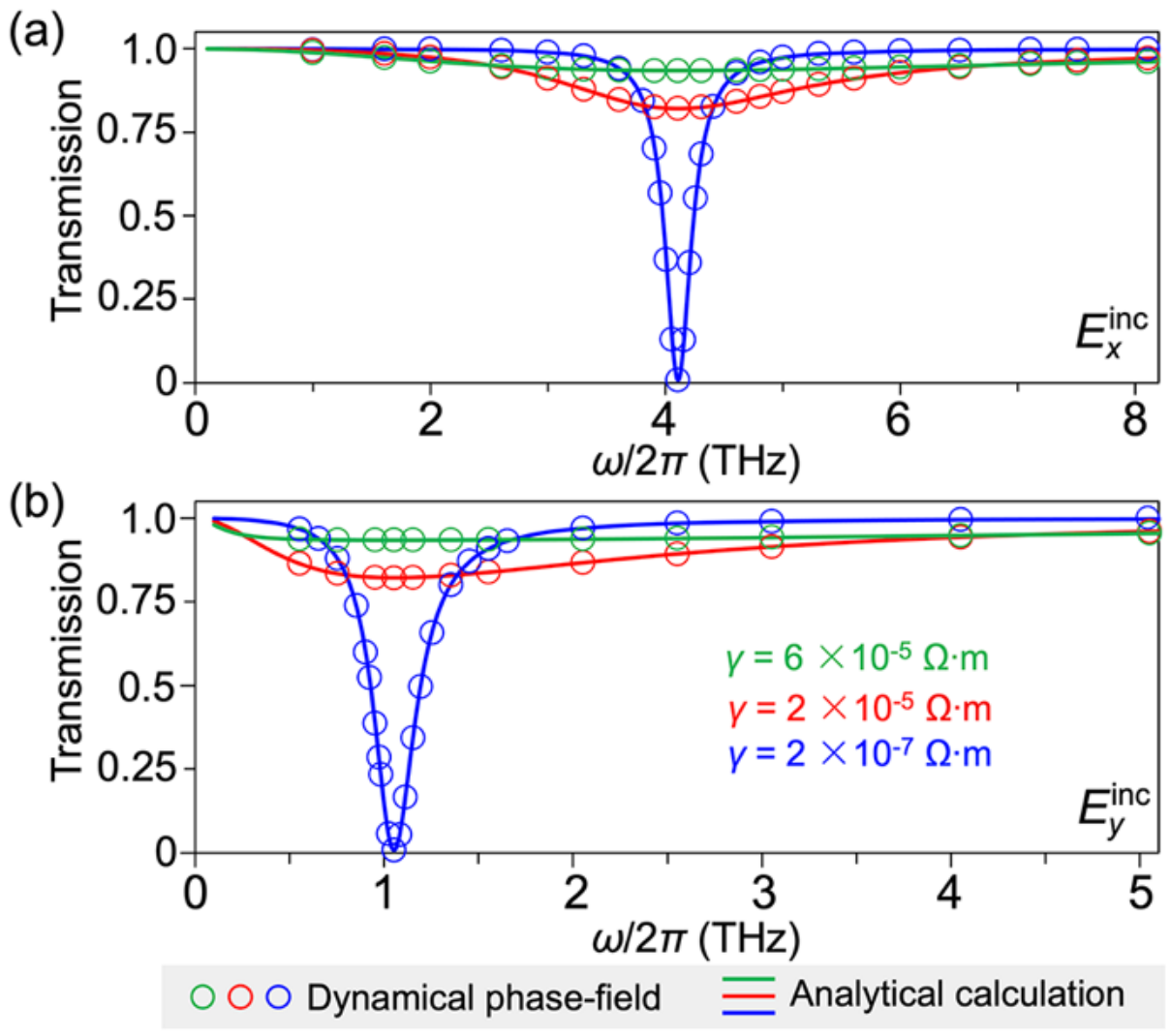

**Figure 3.** Frequency-dependent transmission of (**a**) $E_x^{\mathrm{inc}}$ and (**b**) $E_y^{\mathrm{inc}}$ THz wave across a 10-nm-thick (100) $BaTiO_3$ slab under different damping coefficient $\gamma$.

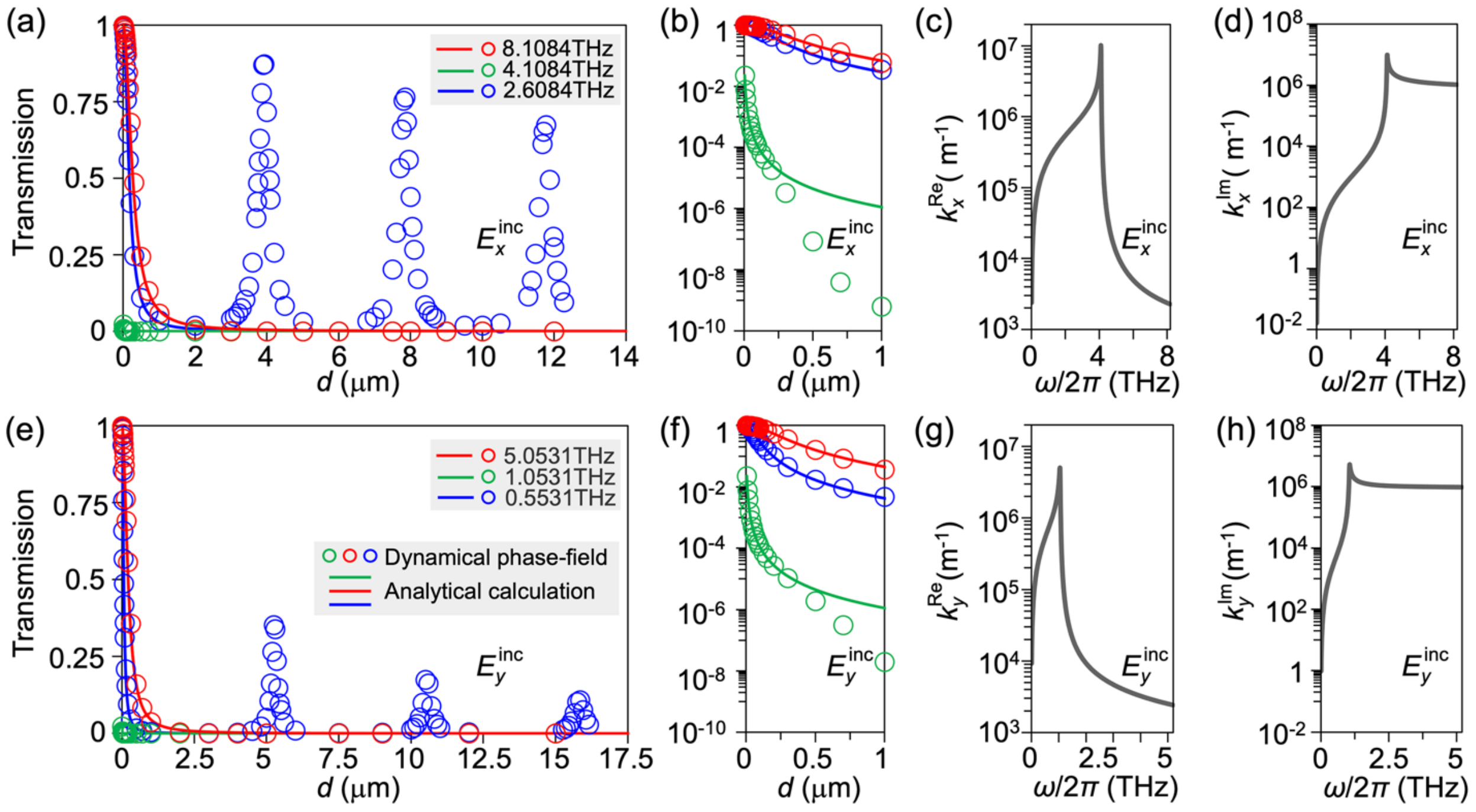

**Figure 4.** Thickness-dependent transmission coefficients for (**a**) $E_x^{\mathrm{inc}}$ and (**c**) $E_y^{\mathrm{inc}}$ across a (100) $BaTiO_3$ slab under a fixed damping coefficient $\gamma$=2×10$^{-7}$ Ω·m at three different frequencies, obtained by analytical calculation (solid lines) and dynamical phase-field simulations (symbols). (**b**,**e**) Their zoom-in figures at low thicknesses, with log-scale transmission coefficients. (**c**,**d**,**g**,**h**) Frequency-dependent real and imaginary part of the complex wavenumber for $E_x^{\mathrm{inc}}$ and $E_y^{\mathrm{inc}}$, obtained by analytical calculations.

### Appendix A: Derivation of the radiation electric field from the oscillating polarization

Let the magnetization **M**, free surface charge density $\rho_{\mathrm{f}}$, and the free charge current density $\mathrm{J_f}$ be zero, the Maxwell's equations in the ferroelectric system can be rewritten as,

$$\nabla\cdot B_i = \nabla\cdot[\mu_0 H_i] = 0, \tag{A1}$$

$$\nabla\cdot D_i = \nabla\cdot(\kappa_0\kappa_{\mathrm{b}}E_i + P_i) = 0, \tag{A2}$$

$$\mu_0\frac{\partial H_i}{\partial t} = -\nabla\times E_i, \tag{A3}$$

$$\kappa_0\kappa_{\mathrm{b}}\frac{\partial E_i}{\partial t} = \nabla\times H_i - \frac{\partial P_i}{\partial t}. \tag{A4}$$

Here $P_i$ refers to the ionic polarization. We use separation of variables and assume **E** and **H** have a plane-wave solution (varying spatially only along $z$) and a harmonic time dependence $\mathrm{e}^{-\mathbf{i}\omega t}$, i.e., $E_i(z,\omega) = E_i(z)\mathrm{e}^{-\mathbf{i}\omega t}$, $H_i(z,\omega) = H_i(z)\mathrm{e}^{-\mathbf{i}\omega t}$, with $i = x, y, z,$ where $\omega$ is the angular frequency. This allows us to rewrite the two time-dependent equations Eqs. (A3-4) in the frequency domain,

$$\mathbf{i}\omega\mu_0 H_i(z,\omega) = \nabla\times E_i(z,\omega), \tag{A5}$$

$$-\mathbf{i}\omega\kappa_0\kappa_{\mathrm{b}}E_i(z,\omega) = \nabla\times H_i(z,\omega) + \mathbf{i}\omega\Delta P_i(z,\omega). \tag{A6}$$

Taking Curls on both sides of Eq. (A5) and substituting the expression of $\nabla\times H_i(z,\omega)$ obtained from Eq. (A6) into Eq. (A5), we have,

$$\nabla(\nabla\cdot E_i) - \nabla^2 E_i = k_{\mathrm{f}}^2 E_i(z,\omega) + \mu_0\omega^2\Delta P_i(z,\omega), \tag{A7}$$

where $k_{\mathrm{f}} = \omega\sqrt{\kappa_0\kappa_{\mathrm{b}}\mu_0}$ and $\Delta P_i(z,\omega) = \kappa_0\chi_{ii}(\omega)E_i(z,\omega)$. It is worth emphasizing that $k_{\mathrm{f}}$ is different from the angular wavenumber of the THz wave in the ferroelectric, $\mathbf{k} = \omega\sqrt{\kappa_0\kappa_{ii}(\omega)\mu_0}$ , where the complex dielectric permittivity $\kappa_{ii}(\omega) = \kappa_{\mathrm{b}} + \chi_{ii}(\omega)$.

Re-arranging Eq. (A2) yields,

$$\nabla\cdot E_i = -\frac{1}{\kappa_0\kappa_{\mathrm{b}}}\nabla\cdot P_i = -\frac{1}{\kappa_0\kappa_{\mathrm{b}}}\left(\frac{\partial P_x}{\partial x} + \frac{\partial P_y}{\partial y} + \frac{\partial P_z}{\partial z}\right). \tag{A8}$$

Under the plane-wave assumption, all physical quantities are spatially uniform within the $xy$ plane, thus $\nabla\cdot E_i = -\frac{1}{\kappa_0\kappa_{\mathrm{b}}}\frac{\partial P_z}{\partial z}$, and $\nabla(\nabla\cdot E_i) = \left(0,0,-\frac{1}{\kappa_0\kappa_{\mathrm{b}}}\frac{\partial^2 P_z}{\partial z^2}\right)$. In this work, we let the initial equilibrium polarization aligns in the film plane and that the incident THz wave does not have a $z$ component. Thus $\nabla\cdot E_i = 0$. Thus, Eq. (A7) can be rewritten as,

$$\frac{\partial^2 E_i(z,\omega)}{\partial z^2} + k_{\mathrm{f}}^2 E_i(z,\omega) = -\mu_0\omega^2\Delta P_i(z,\omega). \tag{A9}$$

Equation (A9) has a solution of,

$$E_i(z,\omega)=\begin{cases} E_i^1(z,\omega)=A\mathrm{e}^{-\mathbf{i}k_\mathrm{a}z}, \quad z<0, \\ E_i^2(z,\omega) = X_1\mathrm{e}^{\mathbf{i}k_\mathrm{f}z}+\dfrac{\mathbf{i}\mu_0\omega^2}{2k_\mathrm{f}}\mathrm{e}^{\mathbf{i}k_\mathrm{f}z}\displaystyle\int_n^z \mathrm{e}^{-\mathbf{i}k_\mathrm{f}z}\Delta P_i(z,\omega)\mathrm{d}z \\ \quad +X_2\mathrm{e}^{-\mathbf{i}k_\mathrm{f}z}-\dfrac{\mathbf{i}\mu_0\omega^2}{2k_\mathrm{f}}\mathrm{e}^{-\mathbf{i}k_\mathrm{f}z}\displaystyle\int_m^z \mathrm{e}^{\mathbf{i}k_\mathrm{f}z}\Delta P_i(z,\omega)\mathrm{d}z, \quad 0<z<d, \\ E_i^3(z,\omega)=B\mathrm{e}^{\mathbf{i}k_\mathrm{a}z}, \quad z>d. \end{cases} \tag{A10}$$

Here $k_\mathrm{a} = \omega\sqrt{\kappa_0\mu_0}$ is the wavenumber of the EM wave in the vacuum; *m* and *n* are two independent constants; $0<z<d$ refers to the ferroelectric layer; $z<0$ and $z>d$ refers to the area below and above the ferroelectric layer, respectively. Equation (A14) indicates the EM wave inside the ferroelectric layer contains both forward-propagating (along +*z*) and backward-propagating (along -*z*) waves as well as the EM wave source at every layer (*i.e.*, at different *z* coordinate) of the ferroelectric. The four unknown coefficients in Eq. (A10) can be obtained by applying the EM wave boundary condition (tangential field continuity and normal flux continuity), i.e., $E_i^1(z=0,\omega) = E_i^2(z=0,\omega)$ ; $E_i^2(z=d,\omega) = E_i^3(z=d,\omega)$ ; $H_i^1(z=0,\omega) = H_i^2(z=0,\omega)$ ; $H_i^2(z=d,\omega) = H_i^3(z=d,\omega)$ , with $i=x,y$. Note that $H_x(z,\omega) = -(\partial E_y/\partial z)/(\mathbf{i}\omega\mu_0)$ and $H_y(z,\omega) = -(\partial E_x/\partial z)/(\mathbf{i}\omega\mu_0)$ . The four coefficients are determined as follows,

$$\mathrm{A} = -\mathbf{i}\omega^2\mu_0\frac{\mathrm{e}^{\mathbf{i}k_\mathrm{f}d}(k_\mathrm{f}-k_\mathrm{a})\int_0^d \mathrm{e}^{-\mathbf{i}k_\mathrm{f}z}\Delta P_i(z,\omega)\mathrm{d}z+\mathrm{e}^{-\mathbf{i}k_\mathrm{f}d}(k_\mathrm{f}+k_\mathrm{a})\int_0^d \mathrm{e}^{\mathbf{i}k_\mathrm{f}z}\Delta P_i(z,\omega)\mathrm{d}z}{\mathrm{e}^{\mathbf{i}k_\mathrm{f}d}(k_\mathrm{f}-k_\mathrm{a})^2-\mathrm{e}^{-\mathbf{i}k_\mathrm{f}d}(k_\mathrm{f}+k_\mathrm{a})^2} \tag{A11a}$$

$$\mathrm{B} = -\mathbf{i}\omega^2\mu_0\mathrm{e}^{-\mathbf{i}k_\mathrm{a}d}\frac{(k_\mathrm{f}-k_\mathrm{a})\int_0^d \mathrm{e}^{\mathbf{i}k_\mathrm{f}z}\Delta P_i(z,\omega)\mathrm{d}z+(k_\mathrm{a}+k_\mathrm{f})\int_0^d \mathrm{e}^{-\mathbf{i}k_\mathrm{f}z}\Delta P_i(z,\omega)\mathrm{d}z}{\mathrm{e}^{\mathbf{i}k_\mathrm{f}d}(k_\mathrm{f}-k_\mathrm{a})^2-\mathrm{e}^{-\mathbf{i}k_\mathrm{f}d}(k_\mathrm{f}+k_\mathrm{a})^2} \tag{A11b}$$

$$X_1 = -\frac{\mathbf{i}\omega^2\mu_0}{2k_\mathrm{f}}\frac{1}{\mathrm{e}^{\mathbf{i}k_\mathrm{f}d}(k_\mathrm{f}-k_\mathrm{a})^2-\mathrm{e}^{-\mathbf{i}k_\mathrm{f}d}(k_\mathrm{f}+k_\mathrm{a})^2}[(k_\mathrm{f}-k_\mathrm{a})^2\mathrm{e}^{\mathbf{i}k_\mathrm{f}d}\int_\mathrm{n}^d \mathrm{e}^{-\mathbf{i}k_\mathrm{f}z}\Delta P_i(z,\omega)\mathrm{d}z$$
$$+(k_\mathrm{f}+k_\mathrm{a})^2\mathrm{e}^{-\mathbf{i}k_\mathrm{f}d}\int_0^n \mathrm{e}^{-\mathbf{i}k_\mathrm{f}z}\Delta P_i(z,\omega)\mathrm{d}z+(k_\mathrm{f}^2-k_\mathrm{a}^2)\mathrm{e}^{-\mathbf{i}k_\mathrm{f}d}\int_0^d \mathrm{e}^{\mathbf{i}k_\mathrm{f}z}\Delta P_i(z,\omega)\mathrm{d}z] \tag{A11c}$$

$$X_2 = -\frac{\mathbf{i}\omega^2\mu_0}{2k_\mathrm{f}}\frac{1}{\mathrm{e}^{\mathbf{i}k_\mathrm{f}d}(k_\mathrm{f}-k_\mathrm{a})^2-\mathrm{e}^{-\mathbf{i}k_\mathrm{f}d}(k_\mathrm{f}+k_\mathrm{a})^2}[(k_\mathrm{f}+k_\mathrm{a})^2\mathrm{e}^{-\mathbf{i}k_\mathrm{f}d}\int_\mathrm{m}^d \mathrm{e}^{\mathbf{i}k_\mathrm{f}z}\Delta P_i(z,\omega)\mathrm{d}z$$
$$+(k_\mathrm{f}-k_\mathrm{a})^2\mathrm{e}^{\mathbf{i}k_\mathrm{f}d}\int_0^m \mathrm{e}^{\mathbf{i}k_\mathrm{f}z}\Delta P_i(z,\omega)\mathrm{d}x+(k_\mathrm{f}^2-k_\mathrm{a}^2)\mathrm{e}^{\mathbf{i}k_\mathrm{f}d}\int_0^d \mathrm{e}^{-\mathbf{i}k_\mathrm{f}z}\Delta P_i(z,\omega)\mathrm{d}z] \tag{A11d}$$

Plugging in the expressions of $X_1$ and $X_2$ into Eq. (A10) leads to elimination of the constants *m* and *n*, yielding an exact solution for the radiation electric field in the ferroelectric film ($0<z<d$), given by,

$$E_i(z,\omega) = -\frac{\mathbf{i}\omega^2\mu_0}{2k_\mathrm{f}[\mathrm{e}^{\mathbf{i}k_\mathrm{f}d}(k_\mathrm{f}-k_\mathrm{a})^2-\mathrm{e}^{-\mathbf{i}k_\mathrm{f}d}(k_\mathrm{f}+k_\mathrm{a})^2]}$$
$$[(k_\mathrm{f}-k_\mathrm{a})^2\mathrm{e}^{\mathbf{i}k_\mathrm{f}d}(\mathrm{e}^{\mathbf{i}k_\mathrm{f}z}\int_z^d \mathrm{e}^{-\mathbf{i}k_\mathrm{f}z}\Delta P_i(z,\omega)\mathrm{d}z+\mathrm{e}^{-\mathbf{i}k_\mathrm{f}z}\int_0^z \mathrm{e}^{\mathbf{i}k_\mathrm{f}z}\Delta P_i(z,\omega)\mathrm{d}z)$$
$$+(k_\mathrm{f}+k_\mathrm{a})^2\mathrm{e}^{-\mathbf{i}k_\mathrm{f}d}(\mathrm{e}^{\mathbf{i}k_\mathrm{f}z}\int_0^z \mathrm{e}^{-\mathbf{i}k_\mathrm{f}z}\Delta P_i(z,\omega)\mathrm{d}z+\mathrm{e}^{-\mathbf{i}k_\mathrm{f}z}\int_z^d \mathrm{e}^{\mathbf{i}k_\mathrm{f}z}\Delta P_i(z,\omega)\mathrm{d}z)$$

$$+(k_\mathrm{f}^2\text{-}k_\mathrm{a}^2)(\mathrm{e}^{-\mathbf{i}k_\mathrm{f}(d-z)}\int_0^d \mathrm{e}^{\mathbf{i}k_\mathrm{f}z}\Delta P_i(z,\omega)\mathrm{d}z+\mathrm{e}^{\mathbf{i}k_\mathrm{f}(d-z)}\int_0^d \mathrm{e}^{-\mathbf{i}k_\mathrm{f}z}\Delta P_i(z,\omega)\mathrm{d}z)]. \tag{A12}$$

One can obtain that $k_\mathrm{f}=1.87\times10^5$ rad/m at $\omega=2\pi\times4$ THz. For $k_\mathrm{f}$ of this order of magnitude, $\mathrm{e}^{\pm\mathbf{i}k_\mathrm{f}z}$ does not change significantly in the range of $0<z<d$ when the thickness $d$ is relatively small (e.g., $10^{-8}\sim10^{-6}$ m). The term $\mathrm{e}^{\pm\mathbf{i}k_\mathrm{f}z}$ can therefore be taken out of the integrand in Eq. (A12). Thus, Eq. (A12) can be reduced to,

$$E_i^\mathrm{rad}(0<z<d,\omega)\approx\frac{\mathbf{i}\omega^2\mu_0\left[\mathrm{e}^{\mathbf{i}k_\mathrm{f}d}(k_\mathrm{f}\text{-}k_\mathrm{a})^2+\mathrm{e}^{-\mathbf{i}k_\mathrm{f}d}(k_\mathrm{f}+k_\mathrm{a})^2+\left(\mathrm{e}^{\mathbf{i}k_\mathrm{f}z}+\mathrm{e}^{-\mathbf{i}k_\mathrm{f}z}\right)\left(k_\mathrm{f}^2\text{-}k_\mathrm{a}^2\right)\right]}{2k_\mathrm{f}[\mathrm{e}^{\mathbf{i}k_\mathrm{f}d}(k_\mathrm{f}\text{-}k_\mathrm{a})^2\text{-}\mathrm{e}^{-\mathbf{i}k_\mathrm{f}d}(k_\mathrm{f}+k_\mathrm{a})^2]}\int_0^d\Delta P_i(z,\omega)\mathrm{d}z\,, \tag{A13}$$

which indicates that the $E_i^\mathrm{rad}$ is largely uniform across the ferroelectric slab even though the excited polarization $\Delta P_i$ is spatially nonuniform. Equivalently, due to the small value of rotation phase $k_\mathrm{f}d$ (~0.187 rad even for $d=1\times10^{-6}$ m), one can use the relation $\mathrm{e}^{\pm\mathbf{i}k_\mathrm{f}d}\approx1\pm\mathbf{i}k_\mathrm{f}d$ to simplify Eq. (A13) into the following,

$$E_i^\mathrm{rad}(0<z<d,\omega)=\frac{\mathbf{i}\omega^2\mu_0(1\text{-}\mathbf{i}dk_\mathrm{a})}{2k_\mathrm{a}\text{-}\mathbf{i}d(k_\mathrm{f}^2+k_\mathrm{a}^2)}\int_0^d\Delta P_i(z,\omega)\mathrm{d}z \tag{A14}$$

If further omitting the terms $\mathbf{i}d(k_\mathrm{f}^2+k_\mathrm{a}^2)$ and $\mathbf{i}dk_\mathrm{a}$ in Eq. (A14) which are much smaller than the other terms under small $d$, and meanwhile assuming $d$ is small enough to have a spatially uniform polarization (thin slab limit), Eq. (A14) can be further simplified to,

$$E_i^\mathrm{rad}(\omega)=\frac{d}{2}\sqrt{\frac{\mu_0}{\kappa_0}}\mathbf{i}\omega\Delta P_i(\omega). \tag{A15}$$

Or equivalently,

$$E_i^\mathrm{rad}(t)=-\frac{1}{2}\frac{d}{\kappa_0 c}\frac{\partial P_i(t)}{\partial t}. \tag{A16}$$

where $c=1/\sqrt{\kappa_0\mu_0}$ is the EM wave velocity in vacuum. We note that Eq. (A16) is the same as expression used in [21]. However, according to the derivations above, Eq. (16) is only applicable to single-polarization-domain, sufficiently thin ferroelectric slab.

## Appendix B: Derivation of the coefficients A$_i$ and B$_i$

Based on the expression of the Landau potential energy density $f^{\mathrm{Landau}}$ of $BaTiO_3$ in [51], ,

$$E_i^{\mathrm{Landau}}=-2\alpha_1 P_i-4\alpha_{11}P_i^3-2\alpha_{12}P_i\left(P_j^2+P_k^2\right)-6\alpha_{111}P_i^5-4\alpha_{112}P_i^3\left(P_j^2+P_k^2\right)-2\alpha_{112}P_i\left(P_j^4+P_k^4\right)-2\alpha_{123}P_iP_j^2P_k^2$$
$$-8\alpha_{1111}P_i^7-2\alpha_{1112}P_i\left[P_j^6+P_k^6+3P_i^4\left(P_j^2+P_k^2\right)\right]-4\alpha_{1122}P_i^3\left(P_j^4+P_k^4\right)-2\alpha_{1123}P_iP_j^2P_k^2(2P_i^2+P_j^2+P_k^2), \quad \text{(B1)}$$

with $i=x,y,z$, and $j\neq i, k\neq i,j$. Note that the summation convention is not used in Eq. (B1). The coefficient $A_i$ can therefore be written as,

$$\mathrm{A}_i=\left.\frac{\partial E_i^{\mathrm{Landau}}}{\partial P_i}\right|_{P=P_i^{\mathrm{eq}}}=-\left.\frac{\partial^2 f^{\mathrm{Landau}}}{\partial {P_i}^2}\right|_{P=P_i^{\mathrm{eq}}}$$

$$=-2\alpha_1-12\alpha_{11}P_i^{\mathrm{eq}2}-2\alpha_{12}\left(P_j^{\mathrm{eq}2}+P_k^{\mathrm{eq}2}\right)-30\alpha_{111}P_i^{\mathrm{eq}4}-12\alpha_{112}P_i^{\mathrm{eq}2}\left(P_j^{\mathrm{eq}2}+P_k^{\mathrm{eq}2}\right)-2\alpha_{112}\left(P_j^{\mathrm{eq}4}+P_k^{\mathrm{eq}4}\right)-2\alpha_{123}P_j^{\mathrm{eq}2}P_k^{\mathrm{eq}2}$$
$$-56\alpha_{1111}P_i^{\mathrm{eq}6}-2\alpha_{1112}\left[P_j^{\mathrm{eq}6}+P_k^{\mathrm{eq}6}+15P_i^{\mathrm{eq}4}\left(P_j^{\mathrm{eq}2}+P_k^{\mathrm{eq}2}\right)\right]$$
$$-12\alpha_{1122}P_i^{\mathrm{eq}2}\left(P_j^{\mathrm{eq}4}+P_k^{\mathrm{eq}4}\right)-2\alpha_{1123}(6P_i^{\mathrm{eq}2}P_j^{\mathrm{eq}2}P_k^{\mathrm{eq}2}+P_j^{\mathrm{eq}4}P_k^{\mathrm{eq}2}+P_j^{\mathrm{eq}2}P_k^{\mathrm{eq}4}), \quad \text{(B2)}$$

For an initial equilibrium polarization of $\left(P_x^{\mathrm{eq}},0,0\right)$, Eq. (B1) reduces to $A_x=-2\alpha_1-12\alpha_{11}P_x^{\mathrm{eq}2}-30\alpha_{111}P_x^{\mathrm{eq}4}-56\alpha_{1111}P_x^{\mathrm{eq}6}$ or $A_y=-2\alpha_1-2\alpha_{12}P_x^{\mathrm{eq}2}-2\alpha_{112}P_x^{\mathrm{eq}4}-2\alpha_{1112}P_x^{\mathrm{eq}6}$. Similarly, based on the expression of the elastic energy density, one can write $E_i^{\mathrm{Elas}}$ as,

$$E_x^{\mathrm{Elas}}=2\left[q_{11}\left(\varepsilon_{xx}-\varepsilon_{xx}^0\right)+q_{12}\left(\varepsilon_{yy}+\varepsilon_{zz}-\varepsilon_{yy}^0-\varepsilon_{zz}^0\right)\right]P_x+2q_{44}\left[\left(\varepsilon_{xy}-\varepsilon_{xy}^0\right)P_y+\left(\varepsilon_{xz}-\varepsilon_{xz}^0\right)P_z\right], \quad \text{(B3}a\text{)}$$

$$E_y^{\mathrm{Elas}}=2\left[q_{11}\left(\varepsilon_{yy}-\varepsilon_{yy}^0\right)+q_{12}\left(\varepsilon_{xx}+\varepsilon_{zz}-\varepsilon_{xx}^0-\varepsilon_{zz}^0\right)\right]P_y+2q_{44}\left[\left(\varepsilon_{xy}-\varepsilon_{xy}^0\right)P_x+\left(\varepsilon_{yz}-\varepsilon_{yz}^0\right)P_z\right], \quad \text{(B3}b\text{)}$$

$$E_z^{\mathrm{Elas}}=2\left[q_{11}\left(\varepsilon_{zz}-\varepsilon_{zz}^0\right)+q_{12}\left(\varepsilon_{xx}+\varepsilon_{yy}-\varepsilon_{xx}^0-\varepsilon_{yy}^0\right)\right]P_z+2q_{44}\left[\left(\varepsilon_{xz}-\varepsilon_{xz}^0\right)P_x+\left(\varepsilon_{yz}-\varepsilon_{yz}^0\right)P_y\right], \quad \text{(B3}c\text{)}$$

where $q_{11}=c_{11}Q_{11}+2c_{12}Q_{12}$, $q_{12}=c_{11}Q_{12}+c_{12}(Q_{11}+Q_{12})$, and $q_{44}=2c_{44}Q_{44}$; $c_{11}$, $c_{12}$, and $c_{44}$ are the elastic stiffness coefficients of the ferroelectric. With the expression of $E_i^{\mathrm{Elas}}$ and plugging in the expressions of $\boldsymbol{\varepsilon}^0$ (see text after Eq.(13)), we can derive,

$$B_i=\left.\frac{\partial E_i^{\mathrm{Elas}}}{\partial P_i}\right|_{P=P_i^{\mathrm{eq}}}=-\left.\frac{\partial^2 f^{\mathrm{Elas}}}{\partial {P_i}^2}\right|_{P=P_i^{\mathrm{eq}}}$$

$$=2\left[q_{11}\left(\varepsilon_{ii}^{\mathrm{eq}}-\varepsilon_{ii}^{0,\mathrm{eq}}\right)+q_{12}\left(\varepsilon_{jj}^{\mathrm{eq}}+\varepsilon_{kk}^{\mathrm{eq}}-\varepsilon_{jj}^{0,\mathrm{eq}}-\varepsilon_{kk}^{0,\mathrm{eq}}\right)\right]-4\left(q_{11}Q_{11}+2q_{12}Q_{12}\right)P_i^{\mathrm{eq}2}-2q_{44}Q_{44}\left(P_j^{\mathrm{eq}2}+P_k^{\mathrm{eq}2}\right), \quad \text{(B3)}$$

If considering an initial equilibrium polarization of $\left(P_x^{\mathrm{eq}},0,0\right)$, one has

$$B_x=2\left[q_{11}\left(\varepsilon_{xx}^{\mathrm{eq}}-\varepsilon_{xx}^{0,\mathrm{eq}}\right)+q_{12}\left(\varepsilon_{yy}^{\mathrm{eq}}+\varepsilon_{zz}^{\mathrm{eq}}-\varepsilon_{yy}^{0,\mathrm{eq}}-\varepsilon_{zz}^{0,\mathrm{eq}}\right)\right]-4\left(q_{11}Q_{11}+2q_{12}Q_{12}\right)P_x^{\mathrm{eq}2} \quad \text{(B4a)}$$

$$B_y=2\left[q_{11}\left(\varepsilon_{yy}^{\mathrm{eq}}-\varepsilon_{yy}^{0,\mathrm{eq}}\right)+q_{12}\left(\varepsilon_{xx}^{\mathrm{eq}}+\varepsilon_{zz}^{\mathrm{eq}}-\varepsilon_{xx}^{0,\mathrm{eq}}-\varepsilon_{zz}^{0,\mathrm{eq}}\right)\right]-2q_{44}Q_{44}P_x^{\mathrm{eq}2} \quad \text{(B4b)}$$

As mentioned in the main text, the total strain at the initial equilibrium state $\boldsymbol{\varepsilon}^{\mathrm{eq}}$ can be obtained by solving the mechanical equilibrium equation different boundary condition. In the present set-up, the ferroelectric slab is stress-free at equilibrium, i.e., $\sigma_{ij}^{\mathrm{eq}}=c_{ijkl}\left(\varepsilon_{ij}^{\mathrm{eq}}-\varepsilon_{ij}^0\right)=0$, thus $\varepsilon_{ij}^{\mathrm{eq}}=\varepsilon_{ij}^0$. As a result, one can write $B_x=-4\left(q_{11}Q_{11}+2q_{12}Q_{12}\right)P_x^{\mathrm{eq}2}$ and $B_y=-2q_{44}Q_{44}P_x^{\mathrm{eq}2}$.

**Appendix C: EM wave transmission across a plain dielectric layer (P=0)**

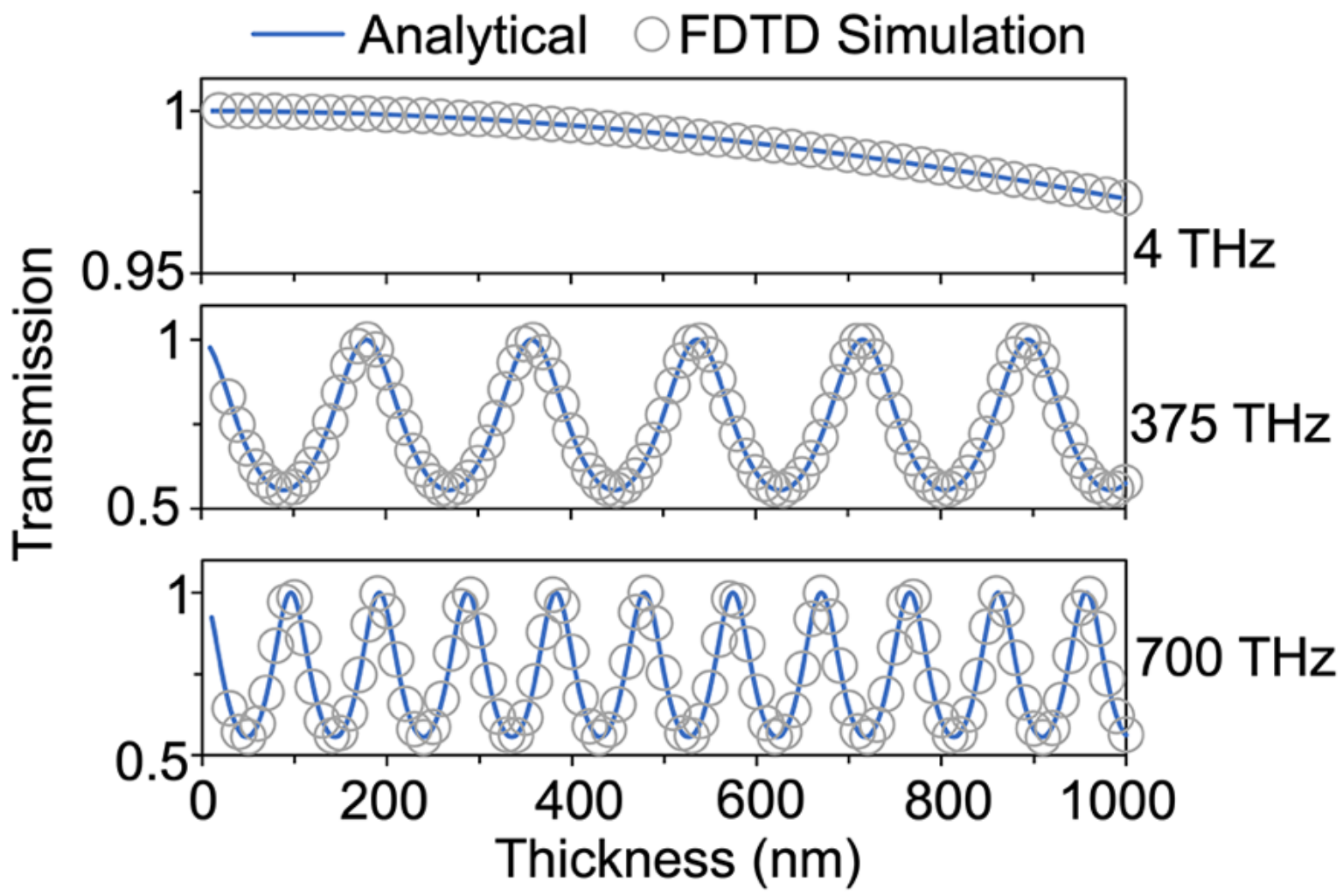

**Figure 5**: Thickness-dependent transmission in dielectric layer ($\kappa_b$=5), obtained from both analytical calculation and FDTD-based numerical simulations. The set-up of the numerical model is the same as Fig. 1a, except that the spontaneous polarization is set as zero (**P**=0). In this case, the transmission is solely determined by the EM wave reflection and refraction at the two ferroelectric/air interfaces and wave interference [33]. Continuous incident EM waves of different frequencies (4 THz, 375 THz, and 700 THz) are considered. The analytical calculation is implemented via the open-source ♯SHAARP package [52].

**Appendix D. Spatial profiles of the standing polarization/THz waves in the $BaTiO_3$**

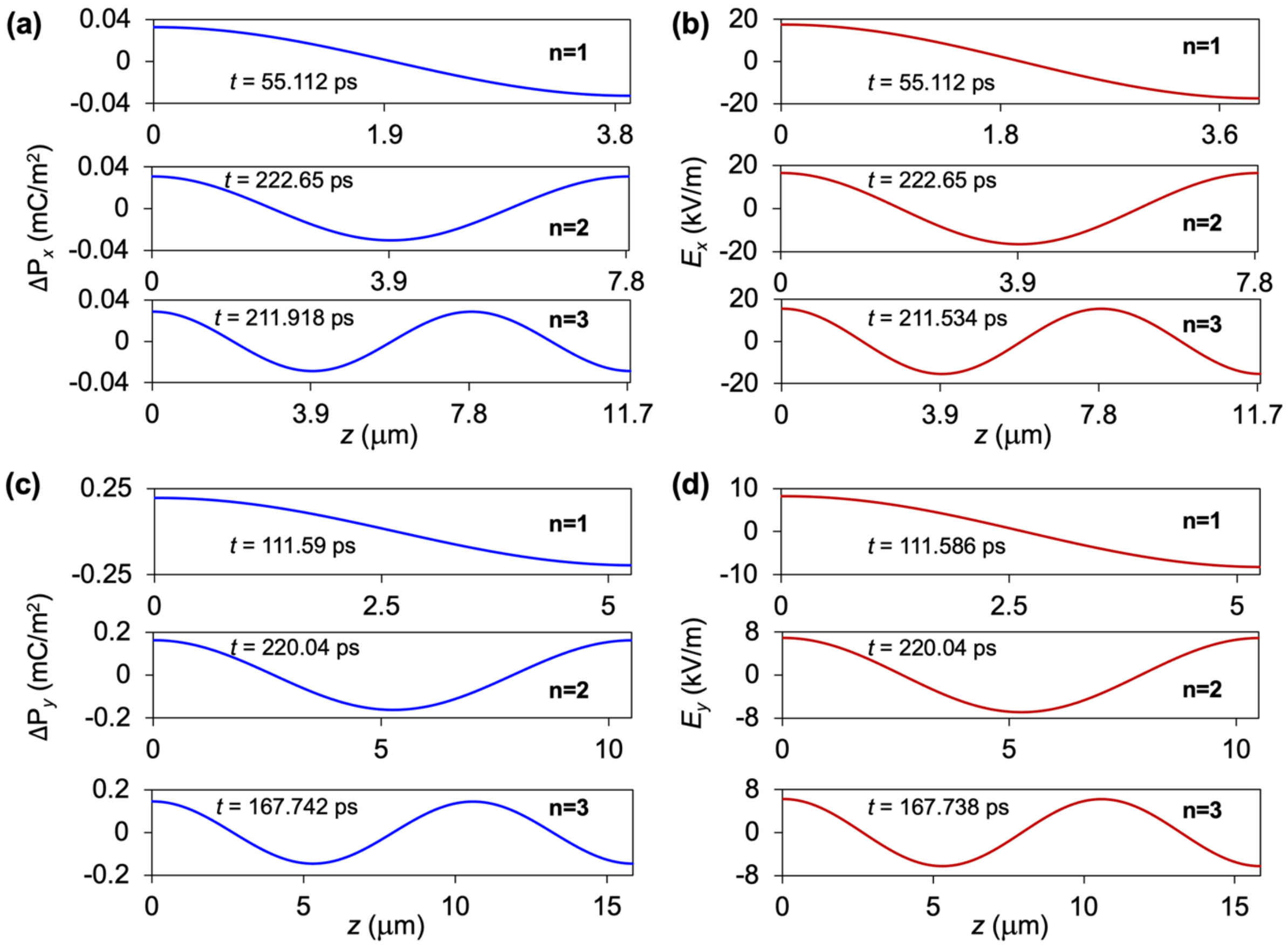

**Figure 6.** Spatial profiles of $n$=1, $n$=2, and $n$=3 mode (**a**) standing polarization wave and (**b**) standing THz wave across the thickness direction ($z$) of the ferroelectric slab for the $E_x^{\mathrm{inc}}$ incident THz wave of 2π×2.6084 THz. The thicknesses of the ferroelectric slab are 3.925 μm, 7.850 μm, and 11.775 μm in the case of $n$=1, $n$=2, and $n$=3 mode, respectively, which result in an identical wavenumber of $\frac{n\pi}{d} = k_x^{\mathrm{Re}}$ of $8.0165\times10^5$ $m^{-1}$. (**c**,**d**) similar data for the $E_y^{\mathrm{inc}}$ incident THz wave of 2π×0.5531 THz. In this case, the thicknesses of the ferroelectric slab are 5.25 μm, 10.5 μm, and 15.850 μm in the case of $n$=1, $n$=2, and $n$=3 mode, respectively, which result in an almost identical wavenumber of $\frac{n\pi}{d} = k_y^{\mathrm{Re}}$ of $5.9556\times10^5$ $m^{-1}$.

## Acknowledgement

The work at University of Wisconsin-Madison was supported by the National Science Foundation (NSF) under the grant number DMR-2237884. The effort at Penn State is supported by the NSF through Grant No. DMR- 2133373. Part of this work was performed under the auspices of the U.S. Department of Energy by Lawrence Livermore National Laboratory under Contract DE-AC52-07NA27344 (B.W.). The dynamical phase-field simulations were performed using Bridges at the Pittsburgh Supercomputing Center through allocation TG-DMR180076 from the Advanced Cyberinfrastructure Coordination Ecosystem: Services & Support (ACCESS) program, which is supported by NSF grants #2138259, #2138286, #2138307, #2137603, and #2138296.